\documentclass[fleqn,usenatbib]{mnras}

\usepackage{newtxtext,newtxmath}

\usepackage[T1]{fontenc}

\DeclareRobustCommand{\VAN}[3]{#2}
\let\VANthebibliography\thebibliography
\def\thebibliography{\DeclareRobustCommand{\VAN}[3]{##3}\VANthebibliography}

\usepackage{graphicx}	
\usepackage{amsmath}	
\usepackage{caption}

\DeclareUnicodeCharacter{2212}{-}
\usepackage{adjustbox}
\usepackage[autostyle]{csquotes}

\title[Galaxies in Abell 426 ]{Morphology, color-magnitude and scaling relations of galaxies in Abell 426}

\author[S. A. Khanday et al.]{
Sheeraz A. Khanday,$^{1}$\thanks{E-mail: sheerazphy@gmail.com}
Kanak Saha$^{2}$,
Nasser Iqbal$^{1}$,
Suraj Dhiwar$^{2,3}$,
Isha Pahwa$^{2}$
\\
$^{1}$University of Kashmir, Hazratbal, Srinagar, J\&K, India\\
$^{2}$Inter-University Center for Astronomy and Astrophysics, Ganeshkhind, Pune-411007, India\\
$^{3}$Department of Physics, Savitribai Phule Pune University, Ganeshkhind, Pune-411007, India\\
}

\date{Accepted XXX. Received YYY; in original form ZZZ}

\pubyear{2021}

\begin{document}
\label{firstpage}
\pagerange{\pageref{firstpage}--\pageref{lastpage}}
\maketitle

\begin{abstract}
We present photometric properties of 183 member galaxies in the Abell~426 cluster using the Sloan Digital Sky Survey (SDSS) imaging and spectroscopic observation. Detailed morphology based on visual classification followed by multi-component image decomposition of 179 galaxies is presented in the SDSS g, r, i-bands. More than 80\% of the members are Early-type galaxies (ETGs), with elliptical, dwarf elliptical (dE), and lenticular morphology and follow the red-sequence in the color-magnitude diagram (CMD). With a few dEs and spirals in the blue cloud, the cluster CMD is nearly unimodal. The dEs are $\sim2$-mag fainter and follow a different Sersic index and central velocity dispersion distribution than their bright counterparts.
Further, we establish the Kormendy relation (KR) and the Fundamental Plane relation (FPR) for 5 different samples of ETGs constructed based on derived physical parameters such as Sersic index, concentration, central velocity dispersion in g, r, i-bands. 
The mean r-band slope and zero-point of the KR are $3.02\pm0.1$ and $18.65\pm0.03$ in close agreement to other cluster ellipticals in the local and higher redshift. Kinematics-based ETG sample produces the least scatter in KR with zero-point getting brighter by $\sim 1.3$~mag from g to i band. The dEs and other low-mass ETGs follow the KR with a similar slope but with $\sim1.3$~mag fainter zero-point and form a parallel KR.The bright ellipticals follow an FPR with $a=1.37\pm0.003$, $b=0.35\pm0.05$ and $c=−9.37\pm0.02$ in the r-band; galaxies tend to deviate from this relation at the low-mass end. 
A catalog with morphology and 2D structural analysis is available online.

\end{abstract}




\section{Introduction}
\label{sec:intro}

Galaxy clusters, being the largest and the massive gravitationally bound systems in the universe \citep{Rines2013,Song2017}, commonly form at the intersection of filaments in the hierarchical structure formation scenario \citep{WhiteRees1978}. Once formed, they grow via continuous accretion of material from the surroundings \citep{KravtsovBorgani2012, Abramoetal2007}. Galaxy clusters, consisting of primarily dark matter, intracluster medium, galaxies and hot gas, are the ideal laboratory for investigating the formation and evolution of galaxies as well as the impact of environment \citep{Sadat1997,Lin2003,Ettori2009,Alonsoetal2020}.

The cluster of galaxies being the densest region, harbours a wide variety of stellar systems including the most massive central galaxy, called the cD galaxy and other galaxy members of almost all Hubble type \citep{Dressler1980}. These galaxies live in a dynamically active environment as they interact with the cluster gravitational potential, with the intracluster medium and with other member galaxies through a variety of physical mechanisms such as tidal interactions and mergers \citep{tt1972,MC1996,barton+2000}, ram pressure stripping~\citep{GunnGott1972,abadi+1999}, strangulation \citep{larson+1980, Peng_2015}. These cluster environment processes are mostly responsible for shaping the fundamental galaxy properties such as their morphology, colour, star formation rate, kinematics \citep{Gregg2017, Book2010, Vollmer2001,Peng_2015,Adhikari2019}. Some of these mechanisms might also cause or be key responsible for galaxies to migrate from the blue cloud to the red sequence and form the so-called color-magnitude relation (CMR) for cluster early-type galaxies \citep{Sandage1972,VisvanathanSandage1977, Boweretal1992}. 

In the nearby universe, a lot of insight on the morphology and evolution of galaxies have been gleaned from studies of the Virgo, Fornax and Coma cluster's galaxy population \citep{godwin+1983,Binggeli1985,Ferguson1989,carter+2008,MichardAndreon2008} but less so from the Abell 426 (hereafter, A426) or Perseus cluster which is another nearby rich cluster of galaxies with a virial mass of the order of $8.5\times 10^{14} {\rm M}_{\odot}$ and at a distance of $\sim 70\,$Mpc \citep{Mathews2006, Wittmann2019}. The Perseus cluster is the brightest X-ray cluster of galaxies in which the X-ray emission from the hot intracluster medium, which dominates the baryonic mass in the cluster, is peaked on the central active galactic nucleus (AGN) of NGC 1275 \citep{edge+1990, Fabianetal2011}. The galaxy population in the core region of the Perseus cluster is dominated by passive, early-type galaxies~\citep{kent-sargent1983s}. Although, it seems to have a dynamically young environment as there are indications of ongoing assembly~\citep{Andreon1994,BrunzendorfMeusinger1999}. These properties have made this cluster to emerge as another unique rich cluster to study the influence of environment on its galaxy population.   

Over the last two decades, there has been a number of studies focused on understanding the nature of galaxy population in this cluster. \cite{BrunzendorfMeusinger1999} presented a homogeneous catalogue of 660 brighter galaxies based on a survey of digitized Schmidt plates taken with Tautenburg 2m telescope. Based on a deeper and higher-resolution imaging studies of a $0.3 \times 0.3 \, {\rm Mpc}^{2}$ central region of the cluster core with WIYN  3.5 m telescope, \cite{Conselice2002,Conselice2003} discovered a sample of 53 dwarf galaxy candidates. More recently, \citet{Wittmann2017} found a population of 89 low surface brightness (LSB) galaxies in the cluster central region based on deep and wide-field survey done with the \textit{William Herschel Telescope}. The cluster environment provides us with an excellent opportunity to probe the formation of the otherwise quiescent dwarf galaxies, especially comparing the physical properties of dwarf ellipticals (dEs) and the bright ellipticals \citep{Kormendy1985}. The question whether dEs are a separate class of galaxies formed differently than the bright ellipticals or dEs are just the fainter version of the bright ellipticals processed in the cluster environment remain still under debate \citep{Kormendy1985,Ferguson1994,Graham_2003,Smith_2004,Matkovi2005,Graham_2005}. Not only the ellipticals, the cluster environment is also known to shape the Lenticular galaxies and their formation \citep{Calvi2012,Cortesietal2011,Barwayetal2011}. However, a systematic study of the galaxy properties from dwarfs to bright ellipticals as well as the Lenticulars and spirals e.g., their quantitative morphology, color magnitude relation and scaling relations such Kormendy \citep{Kormendy1977} and fundamental plane relation \citep{Djorgovski1987} based on a homogeneous sample of galaxies with spectroscopic observation is missing for this cluster.  

In this work, we present quantitative morphology of galaxies  having spectroscopic redshift measurements based on Sloan Digital Sky Survey \citep[SDSS]{Ahn2012}. In order to maintain homogeneity of the data, we restrict our current analysis to SDSS publicly available data only. We perform multi-component 2D modelling of the galaxy light distribution in the SDSS g, r, i passbands. Further, we analyse their color-magnitude relation and study in detail, their scaling relations such Kormendy relation and fundamental plane relation.   

This paper is organised as follows. In Section~\ref{sec:sample}, we present the sample of galaxies and find their cluster membership. In Section~\ref{sec:visual}, we describe their visual morphology. We discuss the GALFIT analysis in Section~\ref{sec:galfit} and the resulting structural parameters in Section~\ref{sec:structure}. The results are presented in Section~\ref{sec:CMR} and \ref{sec:scaling}. Lastly, we discuss and state our primary conclusions in Section~\ref{sec:concl}. Throughout this work, we have used these cosmological parameters: $\Omega_{M} = 0.3$, $\Omega_{\Lambda} =
0.7$, $H_{0} = \,70\rmn{kms}^{-1} \,\rmn{Mpc}^{-1}$. 

\section{Sample selection and Cluster Membership}
\label{sec:sample}
The Perseus cluster or A426 is a rich cluster of galaxies  with a richness of 2 \citep{Abell1958}. The RA (J2000) and Dec (J2000) of the  cluster center is  $03^{\rmn{h}} 19^{\rmn{m}} 47\fs2$, $+41\degr 30\arcmin 47^{\prime \prime}$~\citep{Piffaretti2011}. The redshift $(z)$ of the cluster is $0.01790$~\citep{Struble1999} and the velocity (Helio) is  $5366.28 \, \rmn{km/s}$~\citep{Struble1999}.
The Perseus cluster, though not part of the main surveys of the Sloan Digital Sky Survey \citep[SDSS]{York2000, Stoughton2002}, has been observed as one of the supplementary fields in u, g, r, i and z passbands. For our study, 
we obtain  the photometric and spectroscopic data for galaxies within circular area of  $\sim107$ arcminutes from the cluster center by querying the SDSS database \footnote{https://skyserver.sdss.org/dr12/en/tools/search/sql.aspx}.
The SQL query (see appendix ~\ref{sec:query}) returns a total of 342 galaxies with spectroscopic redshift available in the projected area which corresponds to $\sim 2.4$ Mpc at the redshift of the cluster. We obtain SDSS DR12\citep{Alam2015} images in g, r and i bands. Among the different FITS files available from SDSS, we obtain the \textit{corrected frames} for our analysis. These frames are flat-fielded and bias-subtracted; bad columns and cosmic rays have been interpolated over, sky subtracted, and calibrated to nanomaggies per pixel. The various photometric properties such as Petrosian magnitudes, Petrosian radii, model magnitudes, concentration indices and spectroscopic properties such as redshift and velocity dispersion are used in our analysis. The spectroscopic redshifts ($z$) play a crucial role in establishing the cluster membership of galaxies and the velocity dispersion ($\sigma$) is an important parameter to build the fundamental plane of galaxies.    
  
\begin{figure}
\includegraphics[width=0.45\textwidth]{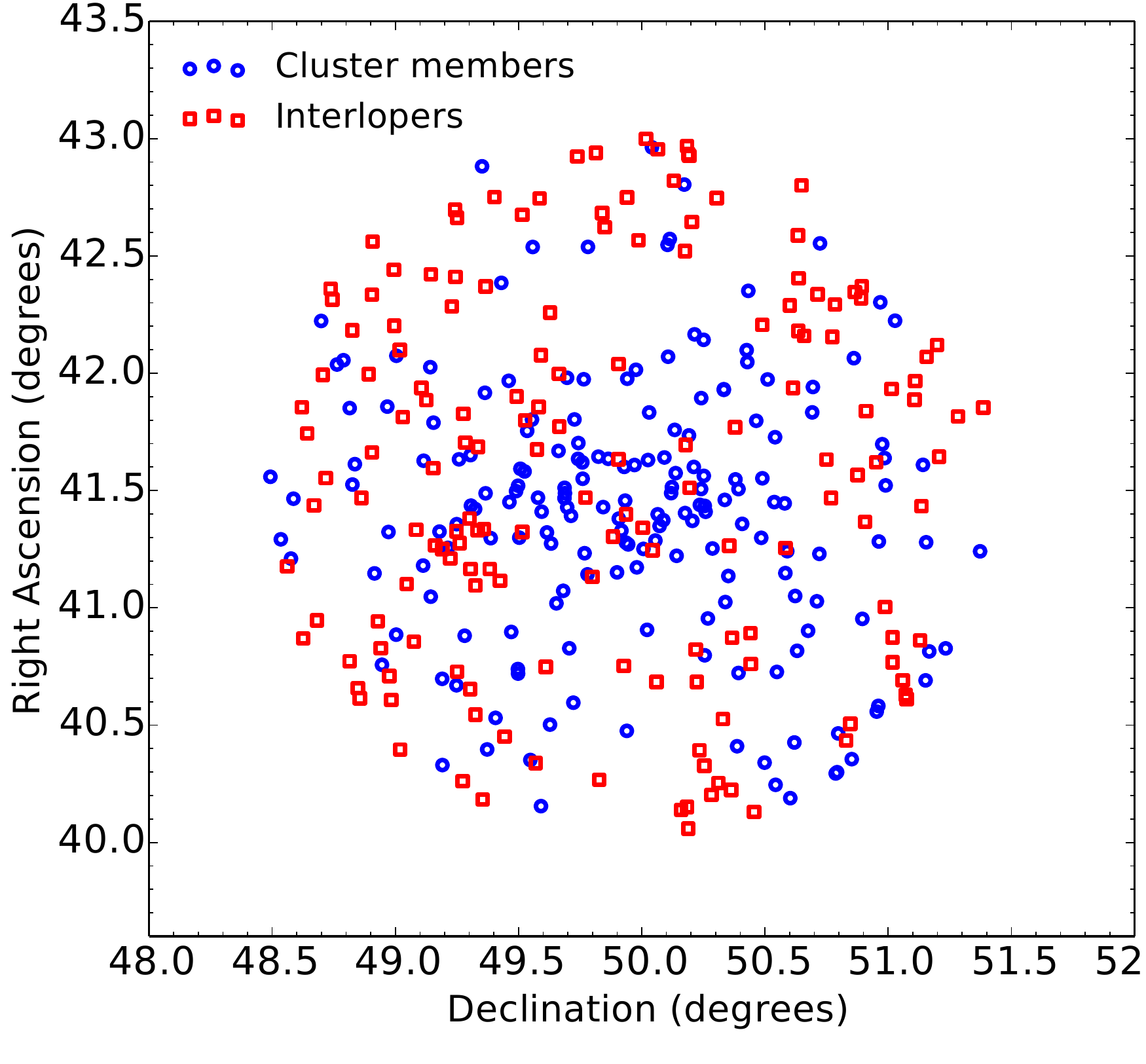}
\caption{Scatter plot of the whole sample of galaxies in the (RA-DEC) plane. The blue circles represent the cluster members and the red squares depict the interlopers.}
\label{fig:radec}
\end{figure}

\begin{figure*}
\includegraphics[width=0.45\textwidth]{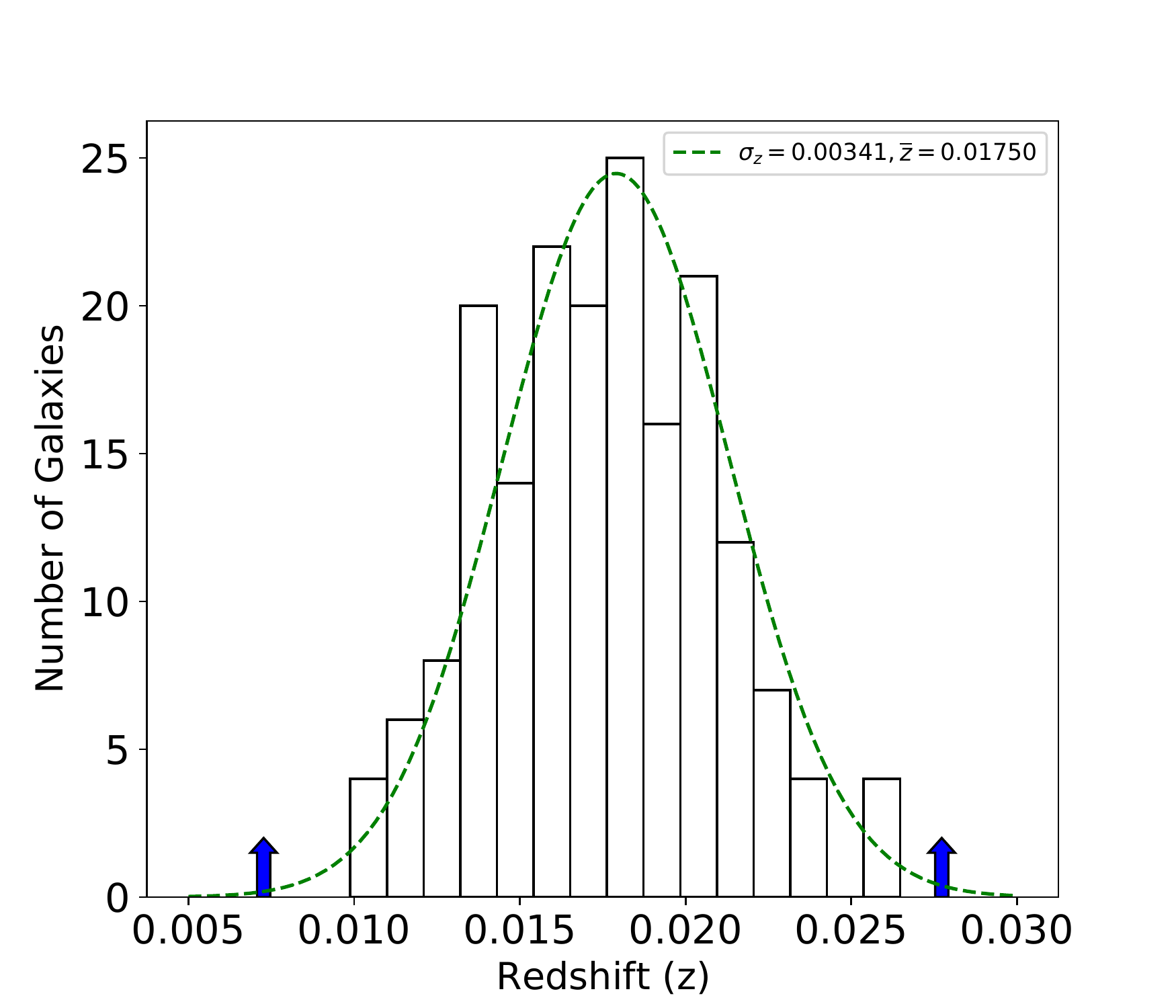}
\includegraphics[width=0.45\textwidth]{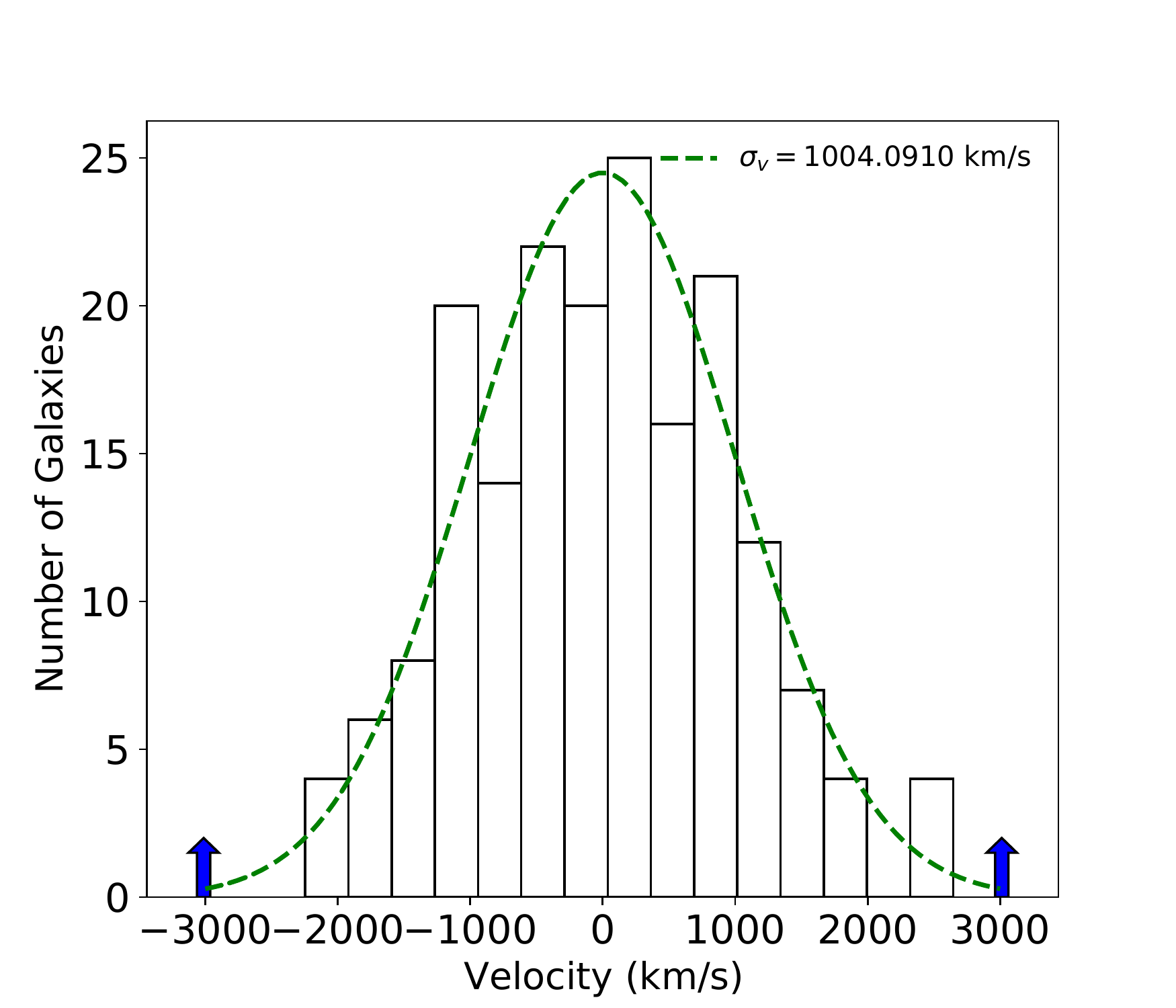}
\caption{The left hand panel shows the redshift distribution and the right hand panel shows the velocity distribution of the galaxies in the sample. The blue arrows denote the $3\sigma$ range about the mean redshift/velocity of the cluster.}
\label{fig:redvel}
\end{figure*}

\par
In the projected area of $107$ arcminutes, the sample of $342$ galaxies has spec-z ranging from $z=0.00988$ to $0.21975$. The search also provides galaxies with negative redshifts which are discarded in the subsequent analysis. A good fraction of these galaxies are interlopers (see Fig.~\ref{fig:radec}) which need to be removed to perform an unbiased dynamical analysis of the cluster based on the cluster members alone \citep{Holmberg1937,Abell1958,TurnerGott1976,Wojtak2007}. We employ direct approach method \citep{YahilVidal1977} based on the calculation of the cluster velocity dispersion ($\sigma_{cl}$) of the galaxy sample and iterative removal of the outliers having velocities higher than $3\sigma$ with respect to the mean velocity. For this, we fit a Gaussian function to the galaxy sample. 

\begin{equation}
    N(v) \, = \, N_{0} \, e^{-\dfrac{(v-v_{0})^2}{2 \, \sigma_{cl}^2}},
    \label{eq:gauss}
\end{equation}

\noindent where $N_{0}$ is the normalization factor; $v_{0}$ and $\sigma_{cl}$ are the mean velocity and the velocity dispersion of the galaxy sample.
We remove all the galaxies with velocities larger than $3\sigma_{cl}$ with respect to the mean and then again fit the Gaussian to the velocity distribution of the remaining  galaxy sample. We repeat this exercise till the mean is saturated and beyond which further iterations do not change mean and sigma of the distribution. Finally, we obtain the velocity dispersion of the cluster as $\sigma_{cl} \sim 1004 ~\rmn{km s^{-1}}$, see the right panel of Fig.~\ref{fig:redvel}. This value of velocity dispersion is in close agreement with previous studies e.g., $\sim 1040 ~\rmn{km s^{-1}}$ by \cite{Aguerri2020}; $\sim 1026 ~\rmn{km s^{-1}}$ by \cite{Girardi_1998}.  
The same procedure has been applied to the spec-z distribution of the galaxies (see the left panel of Fig.~\ref{fig:redvel}). The mean redshift value $\Bar{z}\sim 0.0175$ and the dispersion obtained in redshift is $\sigma_{z} \sim 0.00341$. Within the $3\sigma_{z}$ limit about the mean redshift value, we select $183$ galaxies as the members of the Perseus cluster based on SDSS spectroscopic redshift measurement. In an earlier study by \cite{BrunzendorfMeusinger1999}, they had radial velocities for $\sim 25$\% of their catalogue galaxies. 


\section{Visual Morphology}
\label{sec:visual}
\begin{figure}
\begin{flushleft}
    \includegraphics[width=0.55\textwidth]{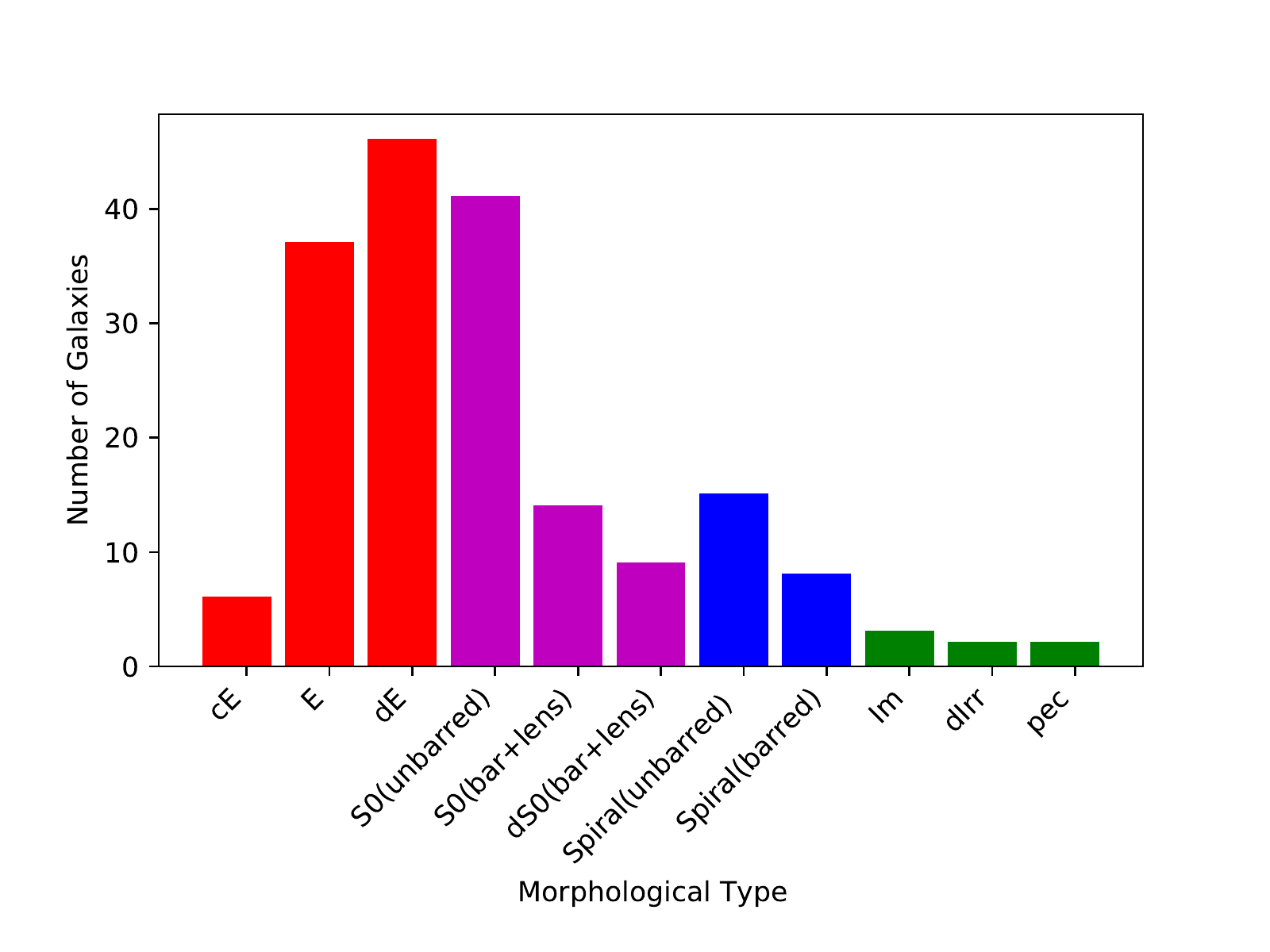}
    \caption{Bar plot of various morphological types of galaxies found in the Perseus cluster. The symbols cE, E, dE and P  denote respectively compact ellipticals, normal ellipticals, dwarf ellipticals and peculiar galaxies. The rest of the symbols have usual meanings. }
    \label{fig:bar_morph}
    \end{flushleft}
\end{figure}

\begin{figure*}
\includegraphics[width =0.8\textwidth]{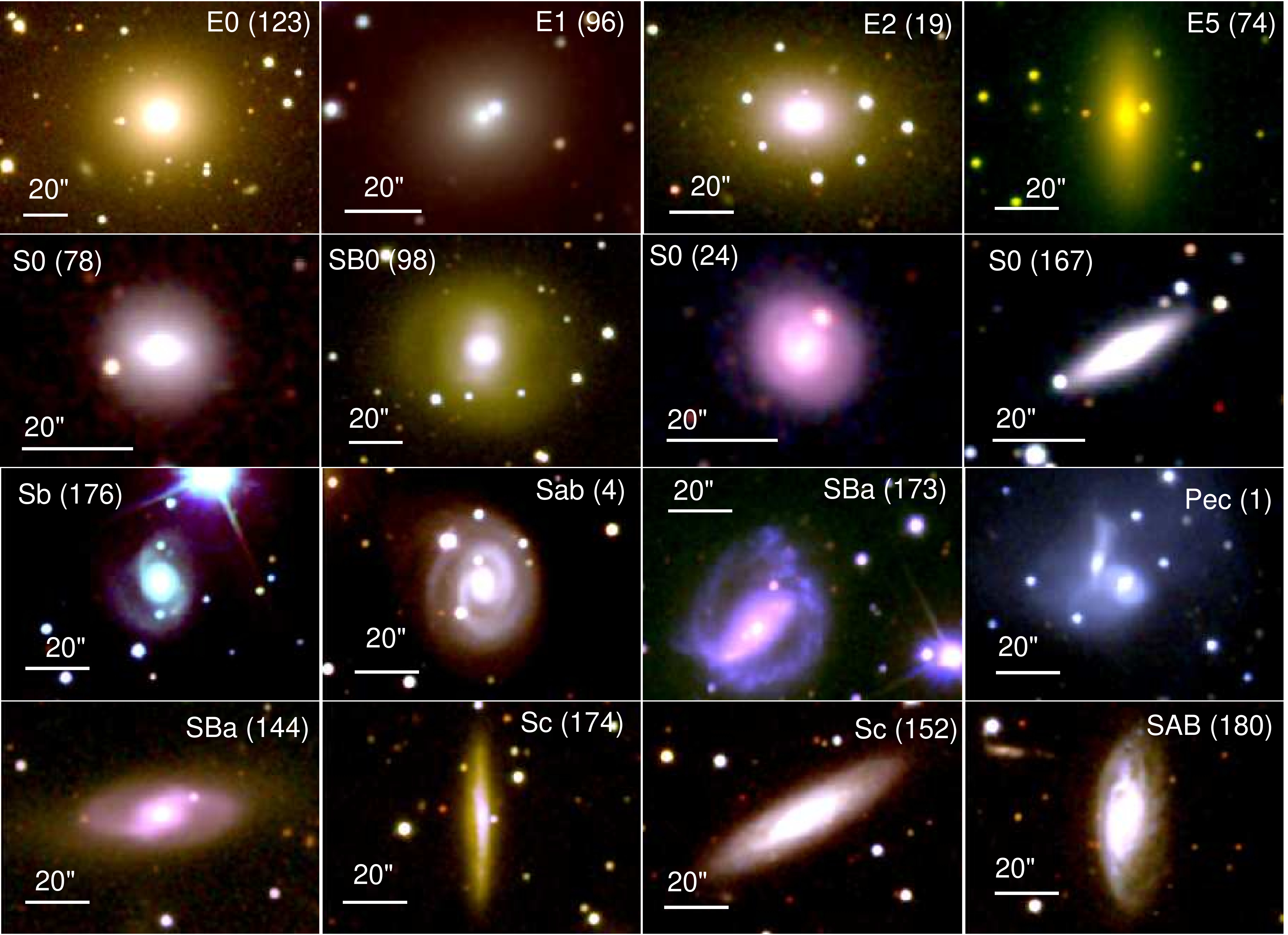}
\caption{First three rows showing the morphology of some of the galaxies from the cluster from early type to late type.  Fourth row showing examples of red spiral galaxies in the cluster. The numbers on the corner are our serial numbers assigned to each galaxy and these also represent the serial number of the galaxies in the online catalog. The horizontal line in each panel represents 20" on the image. }
\label{fig:morph}
\end{figure*}

\begin{figure}
    \centering
    \includegraphics[width=0.5\textwidth]{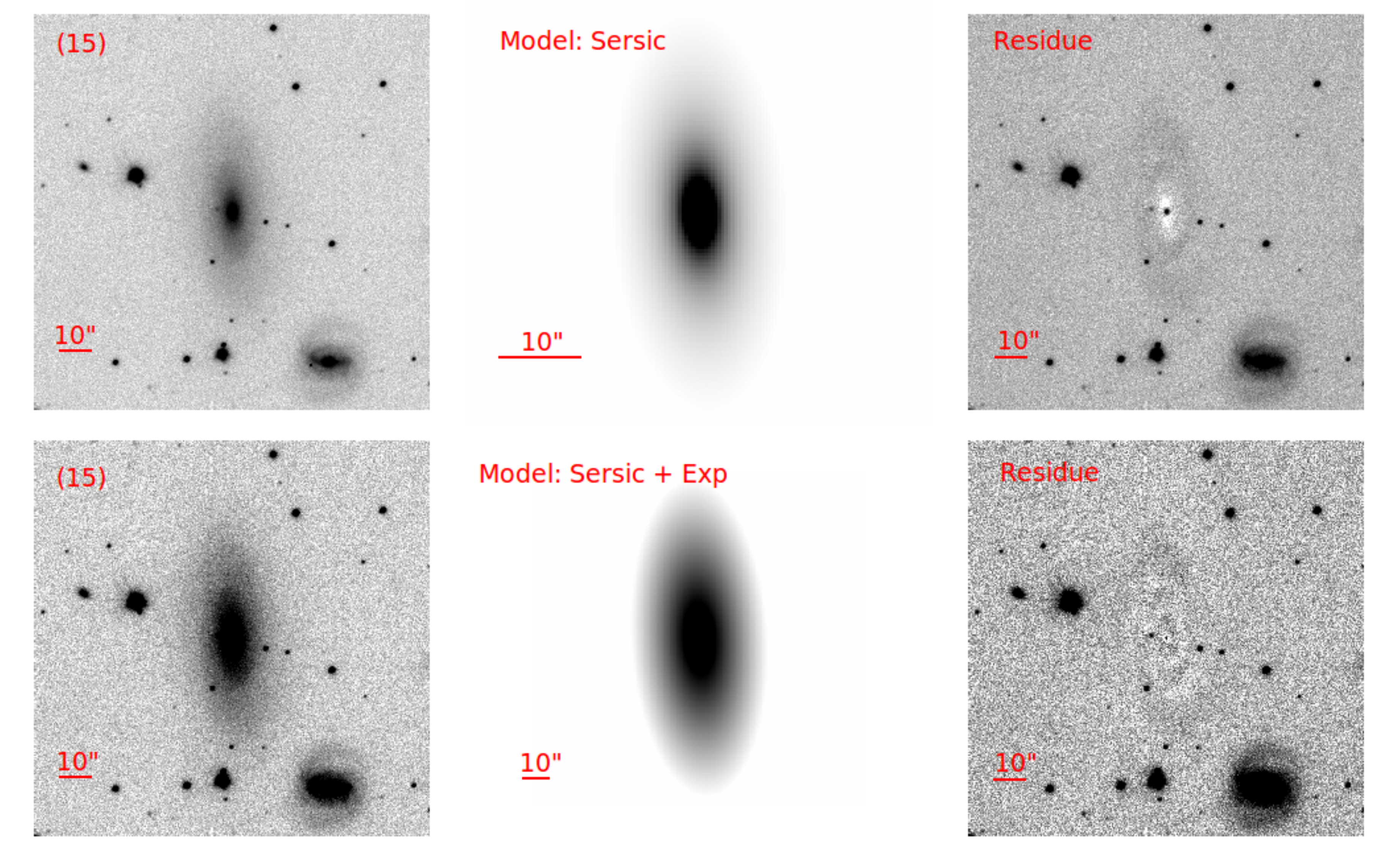}
    \caption{An example of a galaxy where our visual morphology is guided by visual inspection of residual. Based on visual morphology, we expected  galaxy number 15 of our sample to be fitted well with single Sersic. However, as can be seen from the residue in the upper right panel, much light is left in the center. By fitting Sersic plus exponential to the same galaxy, all of the light is taken care of as can be seen from right lower  panel.So we changed its morphological type to S0.}
    \label{fig:2Dvisual}
\end{figure}

\noindent
After confirming the cluster membership of the galaxies based on spectroscopy, we visually examined their morphological features following guidelines provided in \cite{RC31991,Buta1994,Buta2011,Buta2013,Tempel2011,Lintott2008,Lintott2011,Willett2013}. Our classification has been carried out independently by four of the authors in this paper. Classification of edge-on and E6/E5 galaxies (where in En, n=10 (1 - b/a), with b/a being the minor-to-major axis ratio) was ambiguous but we kept the morphology when at least two voted for the same. This worked well for the elliptical, spiral and S0 galaxies but not for dwarf ellipticals (dEs) and compact ellipticals (cE) in our sample. dE and cE classification was done primarily by one of the authors.
We find that majority of the galaxies in the cluster are ellipticals ($\sim 49\%$) and S0s ($\sim 35\%$). We have classified spirals, peculiars and  irregulars in one category and these are around $16\%$ in the total sample. Our visual classification was later guided by 2D modelling using GALFIT \citep{Peng2002} and it's residuals. We have carefully examined the residuals of galaxy images one-by-one before assigning the final morphological type. Our sample is primarily comprised of early-types whose analysis form the major part of this work. The morphological type of each galaxy is presented in the full catalogue. Figure~\ref{fig:bar_morph} shows the bar plot of various morphological types of galaxies found in A426. 
The spiral fraction is low as expected in the high density environment \citep{Dressler1980, Whitmore1992,refId0}. The total number of unbarred and barred spirals in the sample are 15 and 8 respectively. A small fraction of galaxies are either Magellanic irregular (Im) or dwarf-irregular (dIrr) or peculiar (pec) type. Figure~\ref{fig:morph} shows the morphology of some of the early and late type galaxies in the cluster including a few red spirals. The usefulness of examining GALFIT residuals is demonstrated in Figure~\ref{fig:2Dvisual} which shows an instance of a visually classified elliptical turning out to be an S0 galaxy after inspection of the residual. In the following section, we calculate the concentration indices for all morphologically classified sample galaxies. 

\begin{figure*}
\includegraphics[width=0.9\textwidth]{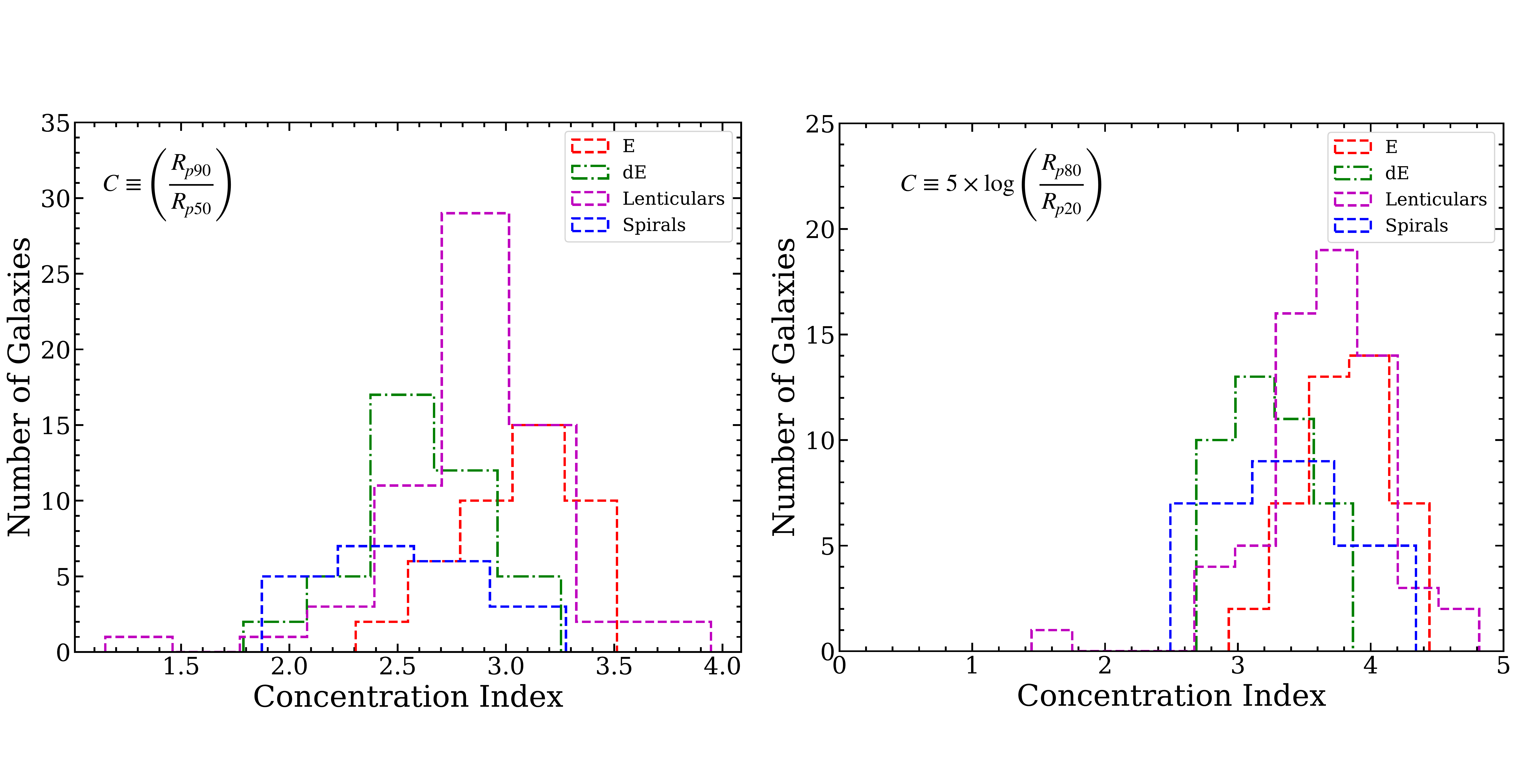}
\vspace{-0.7cm}
\caption{Left panel: histogram showing the concentration indices (based on 90\% and 50\% light) of Es, dEs, S0 and spiral galaxies belonging to the cluster. The median concentration index of Es, dEs, S0s and spirals are 3.10, 2.63, 2.91 and 2.48 respectively. Right panel: Same as in left panel but based on 80\% and 20\% of the total light.}
\label{fig:Chist_new}
\end{figure*}

\begin{figure*}
    \centering
    \includegraphics[width=0.9\textwidth]{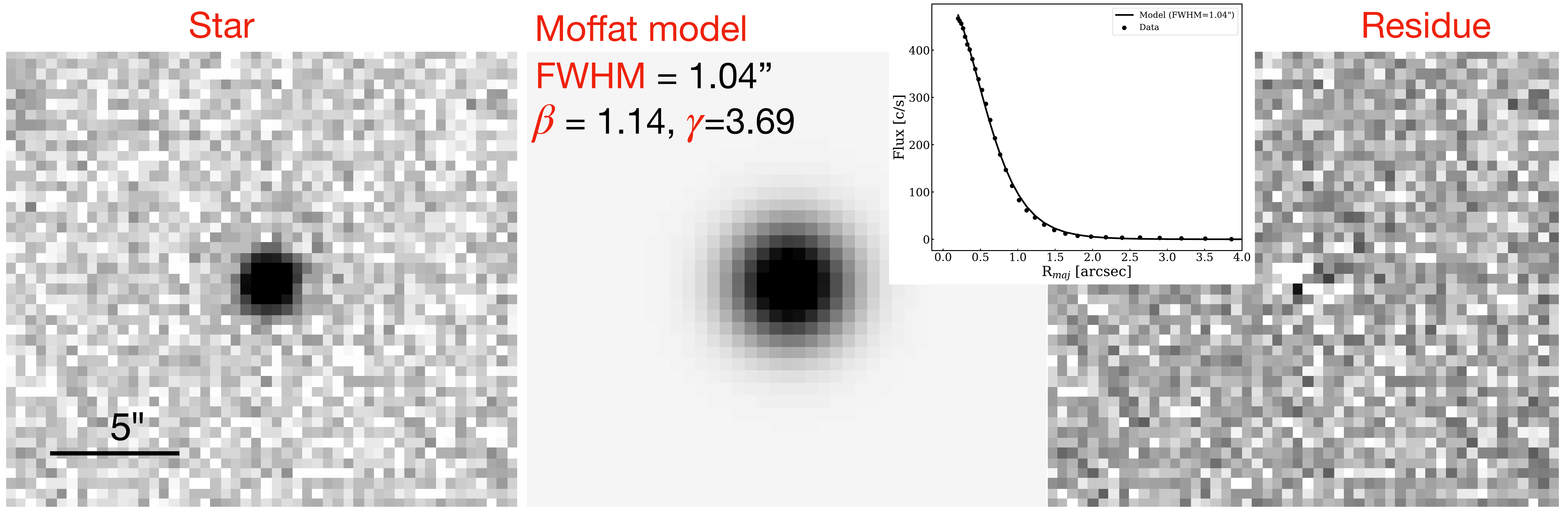}
    \caption{The figure shows  an instance of a star fitted with 2D-Moffat function using GALFIT. The left panel shows the image of a star, the middle panel shows the 2D-Moffat fitted model image and the right panel shows the residue. The inset figure refers to the one-dimensional profile.}
    \label{fig:2DPSF}
\end{figure*}

\subsection{Concentration index}
In the left panel of Fig.~\ref{fig:Chist_new}, we show the histogram of the observed Concentration index defined as \textbf{$C_{95} \equiv R_{p90}/R_{p50}$} \citep{Stratevaetal2001}, where $R_{p90}$ and $R_{p50}$ are the radii in r-band containing 90\% and 50\% of the galaxy's light. Note that we have obtained $R_{p90}$ and $R_{p50}$ from SDSS \textit{PhotoObj} table. For a de Vaucouleurs and a pure exponential profile, the values of \textbf{$C_{95}$} are 5.5 and 2.3 respectively. The steeper the light distribution in the central region, the higher the value of \textbf{$C_{95}$}. Due to the seeing, the observed values are less than these numbers \citep[see for details][]{Stratevaetal2001}. From the distribution of the concentration indices (Fig.~\ref{fig:Chist_new}), it is apparent that most of the galaxies are early-type galaxies (ETGs) in our sample. Following \citep{Stratevaetal2001}, we adopt $C_{95}=2.6$ as the separator for ETGs and late-type galaxies for our spectroscopic sample of galaxies in the cluster. The median values of the concentration indices for the Ellipticals, dwarf ellipticals, Lenticulars and spirals are $3.10$, $2.63$ , $2.91$ and \textbf{$2.48$} respectively in our sample. 
Additionally, we have also calculated the concentration index as $C_{82} =5 \times \log(R_{p80}/R_{p20})$ for each galaxy; where $R_{p80}$ and $R_{p20}$ refer to the radii containing 80\% and 20\% of the total light within $1.5~R_{petro}$ based on the curve of growth analysis \citep{Bershady_2000,Conselice2002}. Following this definition, galaxies having \textbf{$C_{82}>4$} are ellipticals \citep{Conselice2003}. The concentration index histogram for our visually classified galaxies is shown in Fig.~\ref{fig:Chist_new}. The median values of $C_{82}$ for the Ellipticals, dwarf Ellipticals, S0s and spirals are 3.83,3.20, 3.68 and 3.35 respectively. According to this definition, there is a considerable overlap between the Ellipticals and S0s in our sample. The calculation of concentration index reassures our visual morphology classification showing dominant fraction belongs to ellipticals and S0s. We have crossed checked that most of our visually classified ellipticals in our sample have $C_{95} > 2.6$. In the following, we perform the 2D surface photometry for all the galaxies in our sample.

\section{Surface Photometry: 2D decomposition}
\label{sec:galfit}

In order to perform the surface brightness decomposition for our sample of 183 galaxies, we create their cutouts in g, r and i bands from SDSS \textit{corrected frames} containing these galaxies. We consider the cutout size to be $\sim 8$ times the r-band Petrosian radius of the galaxy as provided by SDSS. It is thus ensured that enough of the sky background is included around each galaxy and galaxy outskirts are not near the edge of the frame \citep{Gadotti2009,Kurk2018}. Before proceeding for GALFIT fitting \citep{Peng2002,peng2010}, we remove unwanted sources around our target galaxy by creating masks using the SExtractor \citep{Bertin1996} segmentation map. This is repeated in all three bands and for all galaxies. SDSS-III provides images in nanomaggies. In order for GALFIT to create a good $\sigma$ image, we convert the images from nanomaggy to counts by using the conversion factor from the header of the FITS files.


\subsection{PSF, background and image cutout}
\label{sec:PSF}

To take into account the seeing effects for the image analysis, GALFIT convolves the light profile with  the point spread function (PSF). We model the  PSF for each  galaxy in  all the three bands in the following way. We identify isolated, high signal-to-noise, unsaturated stars in each SDSS frame containing our target galaxies. After examining each star carefully using \textit{imexam} task of IRAF, we extract cutouts of size $20^{\arcsec} \times 20^{\arcsec}$ of five stars (or less) in each frame and median combined them. In doing so, we have ensured that the stars are properly combined such that the brightest pixel of each star falls on top of the others. We then fit the 2D Moffat function to this median combined image and normalized it to obtain the PSF. In this way, we have constructed 106 PSFs (corresponding to the total number of frames containing our sample of  galaxies) in each band. Figure~\ref{fig:2DPSF} shows an example of a PSF used in our sample and the inset figure refers to the one-dimensional profile. The mean FWHM in r-band is $\sim 1^{\arcsec}$, which is nearly equal to the mean FWHM given by SDSS for these galaxies.
We prefer to use model PSF from the Moffat fit over  an empirical one obtained by median combination of the star stamps for two reasons. Firstly, model PSF is smooth and analytic. Secondly, we could compare model PSF FWHM with FWHM from SDSS for each galaxy.


\par
In order to check the effect of image cutout size, we randomly picked 10 galaxies from our sample of ETGs. We made 3 different cutout sizes e.g., $6$, $8$ and $10\times R_{petro}$ for each galaxy, and ran GALFIT on them to examine the changes in the fitting parameters. The last two choices gave nearly identical result. However, the mean difference in magnitude, effective radii, Sersic indices between the first and the last choices turned out to be appreciable, with $\langle\Delta m\rangle \sim 0.04\,\rmn{mag}$, $\langle\Delta  R_{e}\rangle \sim 0.50\,\rmn{pixel}$ and $\langle\Delta n\rangle \sim 0.12$. With smaller cutouts, since enough sky is not included, GALFIT overestimates the fitting parameters. Since $\chi^{2}$ has a smaller value for the last two, we prefer to fix $8\times R_{petro}$ as the cutout size for the rest of the analysis.    
\par
While performing GALFIT analysis for cluster galaxies, it is extremely important to take into account the intracluster light (ICL) or light from the bright nearby sources. We have attempted to include the sky component mimicking the ICL \citep{Fischer2017}. We use the same set up as above but now we estimate the background using SExtractor \citep{Bertin1996} for each cutout. We ran GALFIT in two sets of experiment - in one we use the SExtractor background and in the second, we keep the sky as a free parameter. The mean difference in magnitude, effective radii and Sersic indices are found to be $\langle \Delta m \rangle \sim 0.02\,\rmn{mag}$, $\langle \Delta R_{e} \rangle \sim 0.2\,\rmn{pixel}$ and $\langle \Delta n \rangle \sim 0.06$ respectively. Since differences are not significant, we decided to keep sky component as a free parameter throughout our fitting process, as it gave smaller $\chi^2$ than the first set up. 
We also performed another set of experiments in one of which we add a sky component and in another we do not. It is found that not adding a sky component introduces larger variation in the above parameters. The effect of this is severe in the surface photometry of the dwarf ellipticals in the sample.

\subsection{Single component fitting}
With appropriate image cutout, mask and PSF for each galaxy,  we proceed to perform the two-dimensional modelling using GALFIT that uses the Levenberg-Marquardt algorithm to find the optimum solution to a fit. It models the light distribution in galaxies and other astronomical objects given a set of initial parameters. We use Sersic index n= 4.0 , Petrosian r-band magnitudes and radii, axis ratios and position angles from SDSS as the initial guess parameters for this fit. $\chi^2$ indicates the goodness of a fit but it is always advisable to examine the residue after each fitting. We start with one component Sersic model defined by 
 
 \begin{equation}
 I_{ser}(R) = I_{e} e^{-b_{n}\left[(R/R_e)^{\frac{1}{n}} -1\right]},
 \label{eq:sersic}
 \end{equation}
 
\noindent 
where $I_{e}$ is the surface brightness at the effective radius $R_{e}$, $n$ is the Sersic index controlling the shape of the light profile, and $b_{n}$ is a function of $n$ and is given as $b_n = 1.9992n - 0.3271$ \citep{Graham2005,Ciotti1991} for a wide range of n values. Based on the chi-square and the visual examination of the residuals, $95$ out of $183$ galaxies show good fitting with single Sersic profile. For two of the galaxies in our sample, GALFIT failed to converge . These two galaxies are affected by close neighbouring brighter galaxies. Providing masks did not help us in these two cases. So for them, we have performed simultaneous fitting  of the neighbouring galaxy with the main galaxy. 
 
 \begin{figure}
\begin{flushleft}
\includegraphics[width=0.5\textwidth]{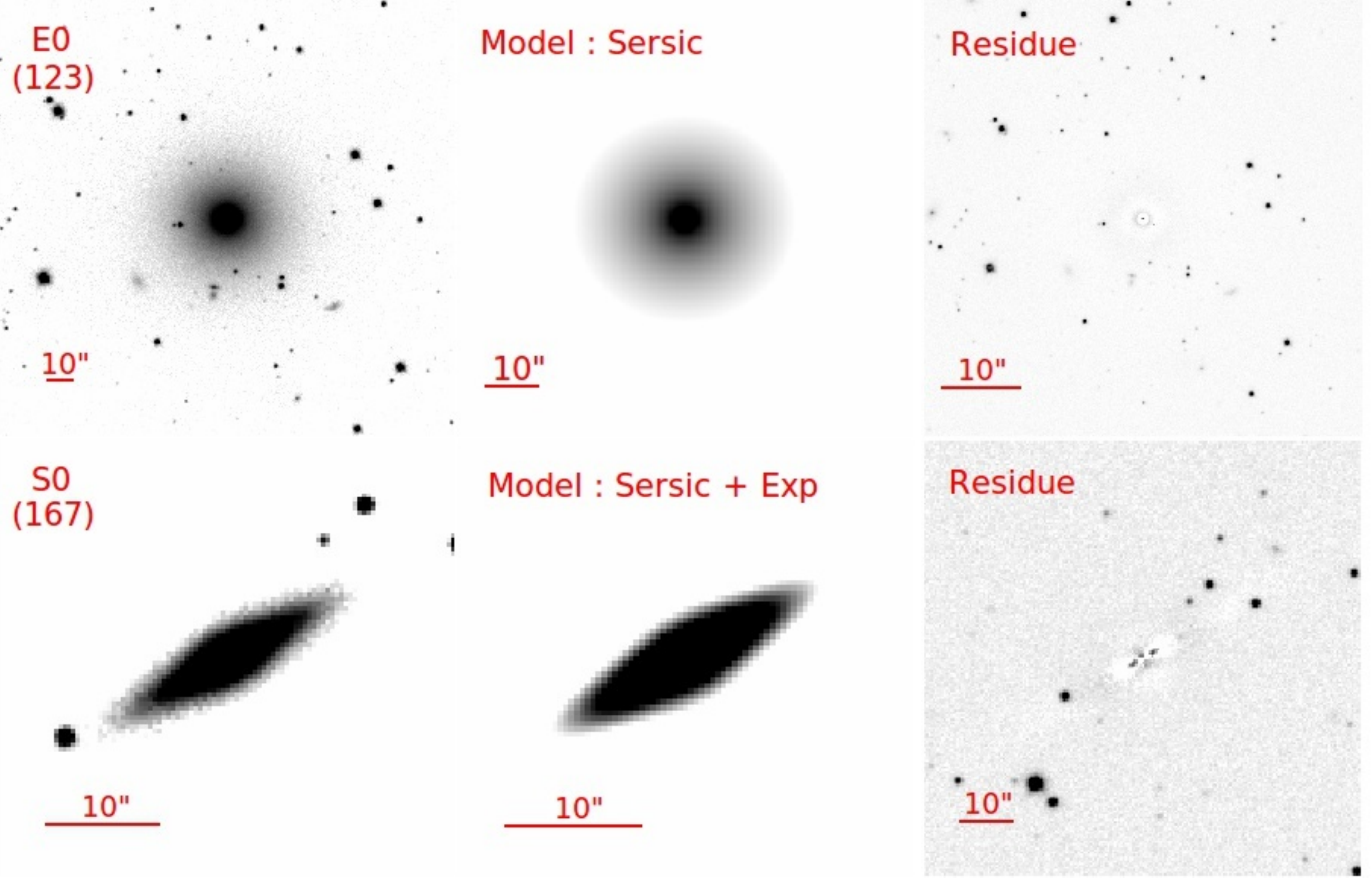}
\includegraphics[width=0.5\textwidth]{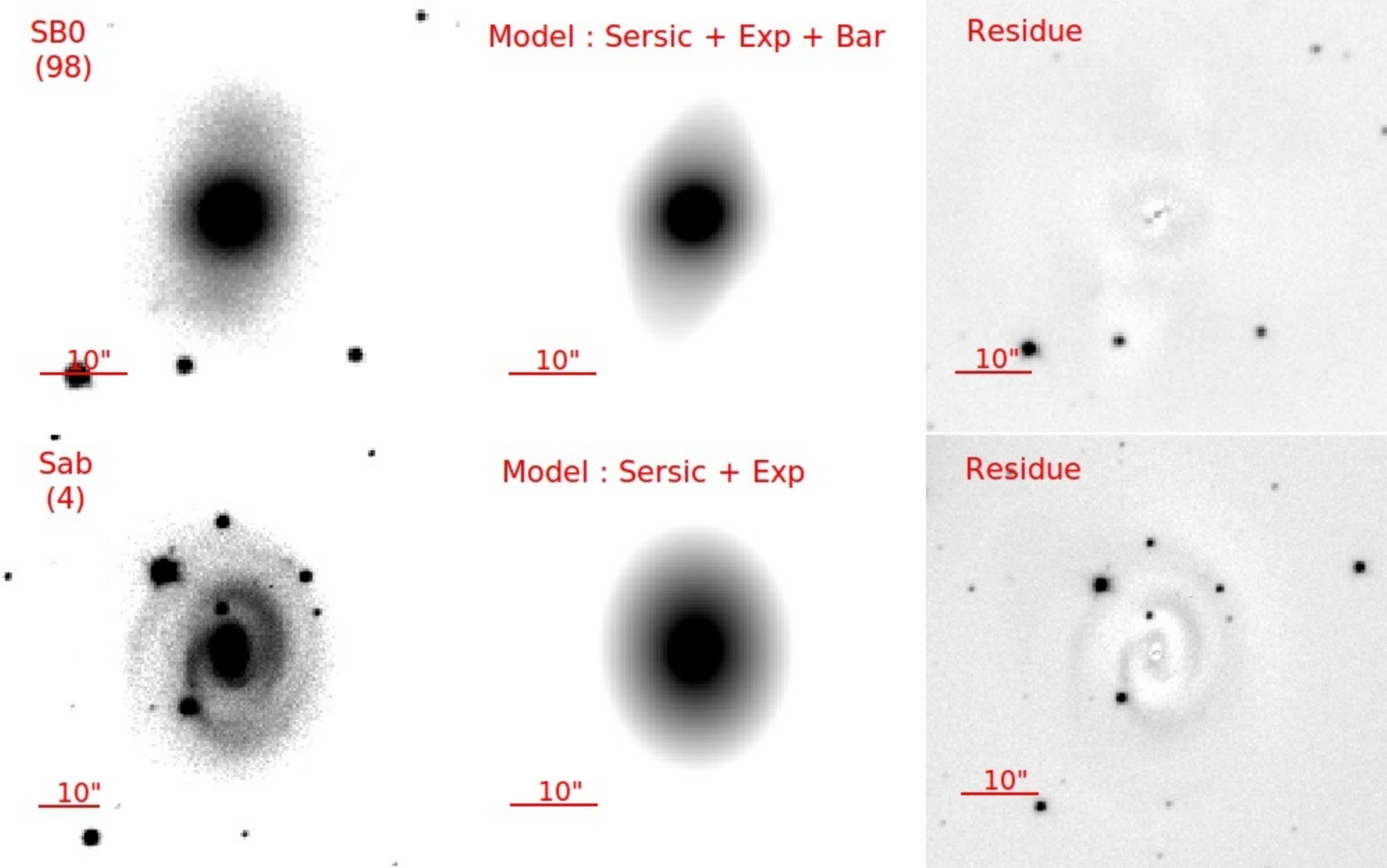}
\caption{GALFIT modelling: A few galaxies are displayed. Real images are in the left panels, model images are in the middle and the residues are in the right panels. All the images are shown in inverted grey scale color and log scale. }
\label{fig:galfit1}
\end{flushleft}
\end{figure}
 
\subsection{Two and three component fitting}
 The galaxies that did not fit well with single Sersic component, showing either large chi-square and / or bad residues,  were subject to two-component fitting. For two-component fitting, we use Sersic plus Exponential profile representing a bulge and a disk. The Exponential disk profile is given by
 
 \begin{equation}
 I_{disc}(R) =I_{0} e^{-R/R_{s}},
 \label{eq:discprofile}
 \end{equation}
 
 \noindent where $I_{0}$ is the disc central surface brightness and $R_{s}$ is the scale length. In two-component fitting, we use the fitted parameters from single Sersic as the initial guess for the bulge component. For fitting disk component, we used brighter magnitude($\sim 1mag$) and larger value of scale radius ($\sim 2 r_{e}$ of the single Sersic fitting) compared to the bulge as the initial starting parameters. In most cases, we have used value around bulge position angle as the initial guess for the disk position angle. However, in some of the cases, where it was not so obvious or there was shift between bulge and disk position angles, the exponential fit position angle (expPhi\_{r}) from SDSS was utilized as the initial guess. In total, 76 galaxies were fitted well with two components.
 
\noindent Once again guided by the residues and chi square, the remaining galaxies, mostly the strongly barred ones, were subject to three component fitting. We add Sersic profile as the third component to account for the bar in each of these galaxies\citep{Kurk2018}.  Except for the bulge Sersic index, we kept all the fitted parameters from the two-component fitting as the initial guesses for fitting bulge and disk in three component fitting. For the bulge, we used Sersic index, $n_{bulge} = 2$ as the initial guess value. For bar component, we used bar magnitude fainter ($\sim 1 mag$) than the disk, but brighter than the bulge  in two component
fitting, effective radius as the 60\% of the disk in two component fitting, n = 0.7, b/a = 0.5 as the initial guesses. 9 galaxies were fitted with three component fitting.
 
To summarize, we fit $95$ galaxies with single Sersic, $75$ galaxies with Sersic plus Exponential and $9$ galaxies with double Sersic plus Exponential profile. We could not fit well $4$ galaxies with any combination of the profiles. Two of them are merging systems; third one is crowded with bright stars that GALFIT fails to converge to any solution and fourth one has a prominent dust lane. We have kept these four galaxies out of our further analysis.

\par
After fitting all the galaxies in $r$-band, we follow the same steps to fit the galaxies in $g$- and $i$-bands. We use the results of $r$-band decomposition as initial guess in fitting galaxies in $g$- and $i$-bands. Figure~\ref{fig:galfit1} displays some of the fitting and their residuals in r-band.   

\noindent Once all the fittings are completed, we estimate their absolute Magnitudes using the following relation

\begin{equation}
     M_{\lambda} = m - 5(\log(D_L)-1) - K_{\lambda} - A_{\lambda},
     \label{eq:absolutemag}
\end{equation}

where $m$ denotes the model magnitude from our fitting, D$_{L}$ is the luminosity distance, estimated using the spec-z,  K$_{\lambda}$ is the K-correction following \cite{Omilletal2011}, and A$_{\lambda}$ is the extinction correction \citep{SchlaflyFinkbeiner2011}, for which we use the dust maps from NASA//IPAC INFRARED SCIENCE ARCHIVE{\footnote{https://irsa.ipac.caltech.edu/applications/DUST/}}. All magnitudes henceforth are foreground dust and K-corrected.

\begin{figure}
\includegraphics[width=0.45\textwidth]{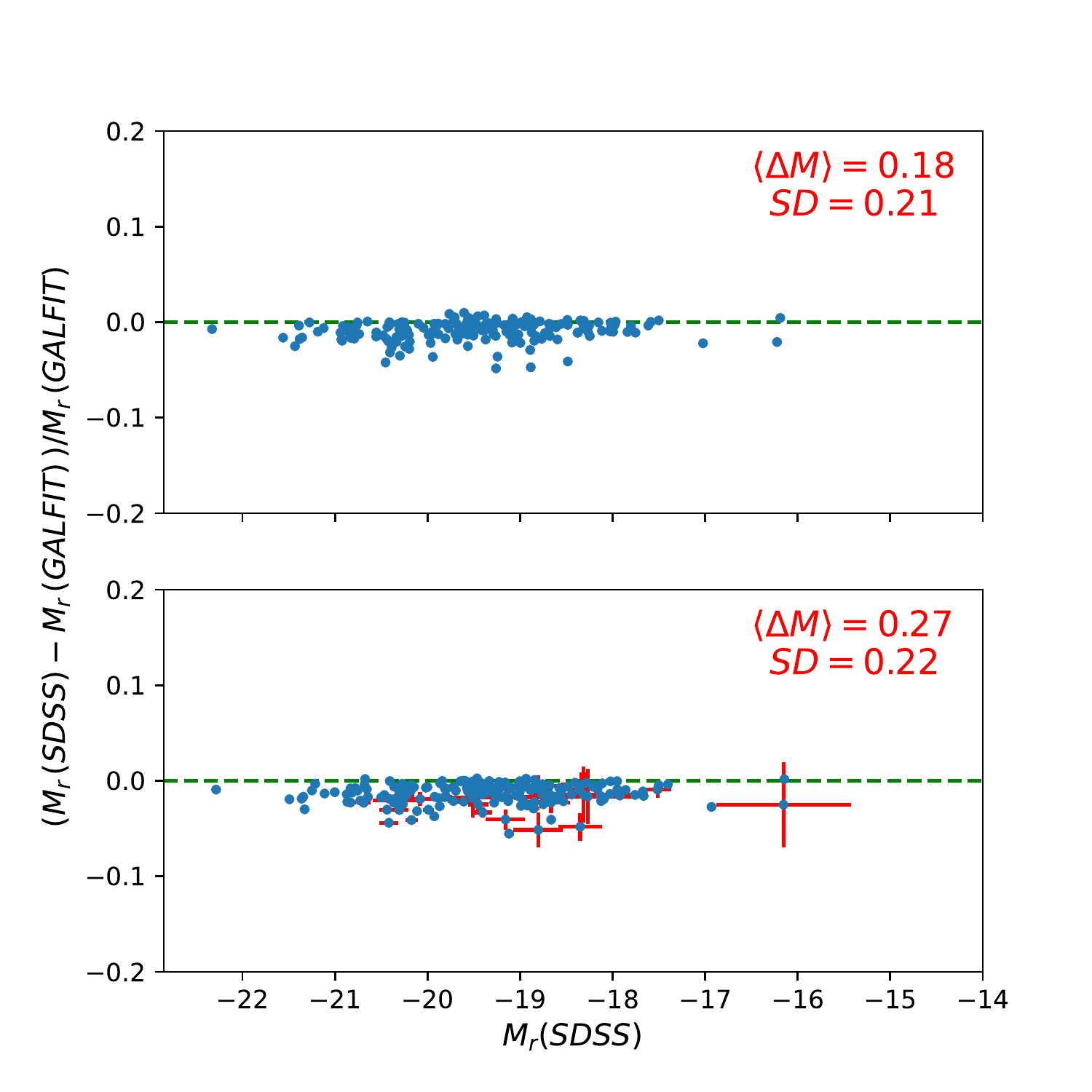}
\caption{The plot showing the relative difference between the SDSS and the GALFIT absolute magnitudes against SDSS absolute magnitudes in $r$-band. In the upper panel,  Composite Model Magnitudes while in the lower panel,  Petrosian magnitudes from SDSS are used for comparison. Here ($\langle \Delta M\rangle$) represents the mean difference of absolute magnitudes and SD represents standard deviation.  }
\label{fig:mag_comp_wSDSS}
\end{figure}

Based on the mean difference between the GALFIT and SDSS magnitudes ($\langle \Delta M\rangle$), we find that our estimated GALFIT magnitudes agree better (see Figure~\ref{fig:mag_comp_wSDSS}) with the SDSS cmodel magnitudes (upper panel) than with the Petrosian magnitudes (lower panel) although systematic difference persists as seen in $g$ and $i$ bands in Figure ~\ref{fig:gidis} in the appendix~\ref{sec:Abs_gi}. Of the possible factors that could have contributed to the observed systematic differences are: neighbouring contamination, improper sky subtraction. For this, we picked randomly a sample of 20 galaxies and proceed to fit with or without adding a sky component in the GALFIT and found the mean difference in magnitude to be $\sim 0.08$~mag. The systematic could also be due to the way SDSS calculates its magnitudes (see appendix~\ref{sec:Abs_gi}). In our case, we allow model profile to fit the data without any truncation.  

\section{Structural Parameters}
\label{sec:structure}

All ellipticals (E+dE) in our sample are well modelled by a single Sersic profile while the lenticulars and spirals are fitted either by two or three components e.g., Sersic plus Exponential or double Sersic plus Exponential depending on their morphology. We have visually examined each GALFIT residuals before finalizing their best fit parameters. We briefly discuss here the overall fitting parameters of the early-type galaxies.

\subsection{Ellipticals, Dwarf ellipticals and Compact ellipticals}
\label{sec:Es}

Altogether, we have 84 ellipticals (E+dE) in our sample and if we include the four dwarf spheroidals and one edgeon dwarf, they constitute about $\sim\,49\%$ of our sample. Our findings are in good agreement with previous studies \citep{Dressler1980} as well as with nearby Virgo cluster in which there are about $\sim\, 48\%$ E+dE galaxies \citep{Binggeli1985,MichardAndreon2008}. However, the E+dE fraction is slightly on the downside, $\sim\, 35\%$ in Coma cluster \citep{MichardAndreon2008}. Based on our quantitative morphological classification, there are 43 ($23$\%) bright elliptical galaxies (E), spanning a wide range in their ellipticity from E0 to E5. There are 41 dwarf ellipticals  i.e., $22$\% in our sample. This is somewhat lower than the value of dE $\sim 27\%$ found in Coma \citep{MichardAndreon2008}. The fraction of dEs in Virgo cluster $\sim 44\%$ -- twice our value\citep{Binggeli1985, MichardAndreon2008}.

\begin{figure}
\includegraphics[width=0.45\textwidth]{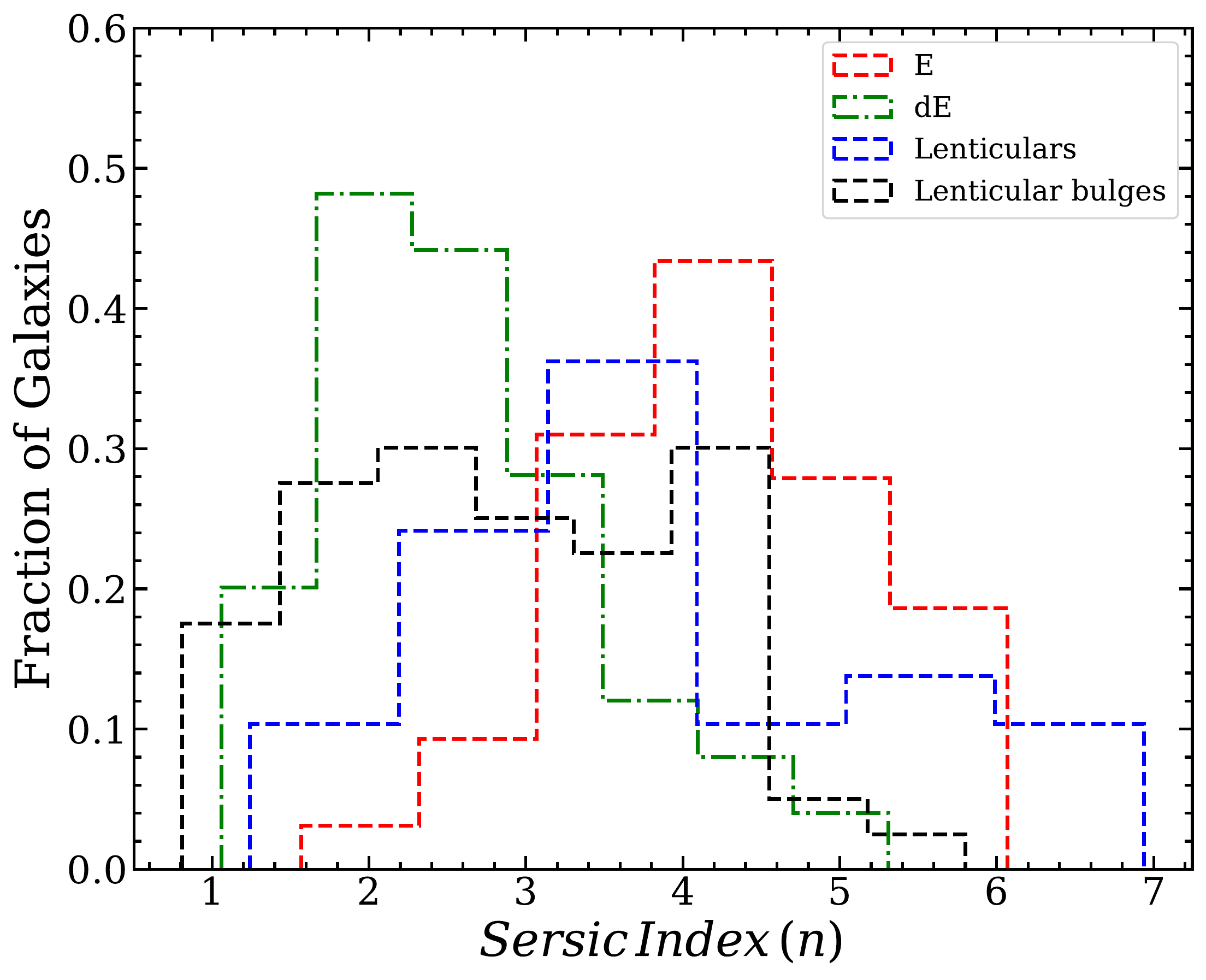}
\includegraphics[width=0.45\textwidth]{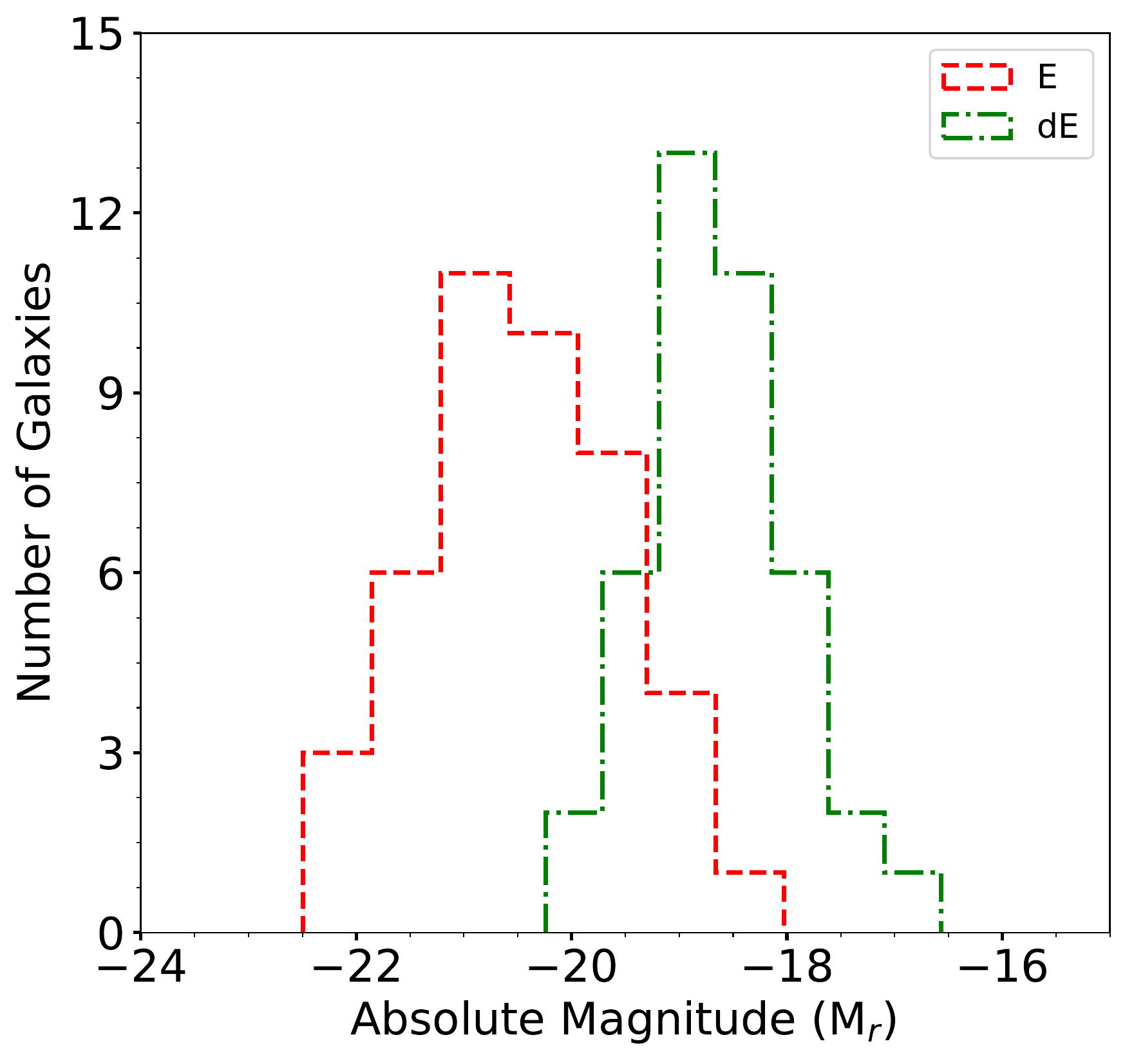}
\caption{Upper panel: histogram of Sersic indices in $r$-band for the Es, dEs, S0s (single Sersic fitting) and bulges of S0s (with 2 or 3 component fitting, black dashed line). The median Sersic indices of Es, dEs, S0s and bulges of S0s are 4.1, 2.5, 3.79 and 2.77 respectively. Bottom panel: histogram of absolute magnitudes in $r$-band for the Es and dEs. The median absolute magnitudes of Es and dEs are $M_r = -20.5$ and $M_r = -18.7$ respectively.}
\label{fig:struc_param}
\end{figure}

The upper panel of Fig.~\ref{fig:struc_param} shows the distribution of the Sersic indices ($n$) for ellipticals and  dwarf ellipticals. Clearly, dEs follow a different Sersic distribution than the bright ellipticals. The median Sersic index for the bright ellipticals is $n_{med} =4.1$ while it is $2.5$ for the dEs --indicating that the dEs are not centrally as bright as the ellipticals. The bottom panel of Fig.~\ref{fig:struc_param} displays their absolute magnitude distribution in the r-band ($M_r$). Evidently, the dEs follow a different luminosity distribution with a median value of $-18.7$~mag which is $\sim 2$~mag fainter than their brighter counterpart having a median value of $-20.5$~mag. Note that our sample of ellipticals (E+dEs) cover a wide range in their absolute magnitude from $\sim -16.6$ to $-22.5$. The effective radii of the bright ellipticals $R_e$ lie in the range ($0.5 - 8.3$)~kpc at the mean redshift of the cluster while for the dwarf ellipticals, effective radii lie in the range of ($0.4 - 6.1$)~kpc. To summarize, we find that the dEs are fainter, less concentrated and smaller than the bright ellipticals. It might be tempting to brand the dEs as the fainter version of the bright ellipticals \citep{Graham_2005} but one must wait to gather full kinematics to conclude.  

\begin{figure*}
\includegraphics[width=0.42\textwidth]{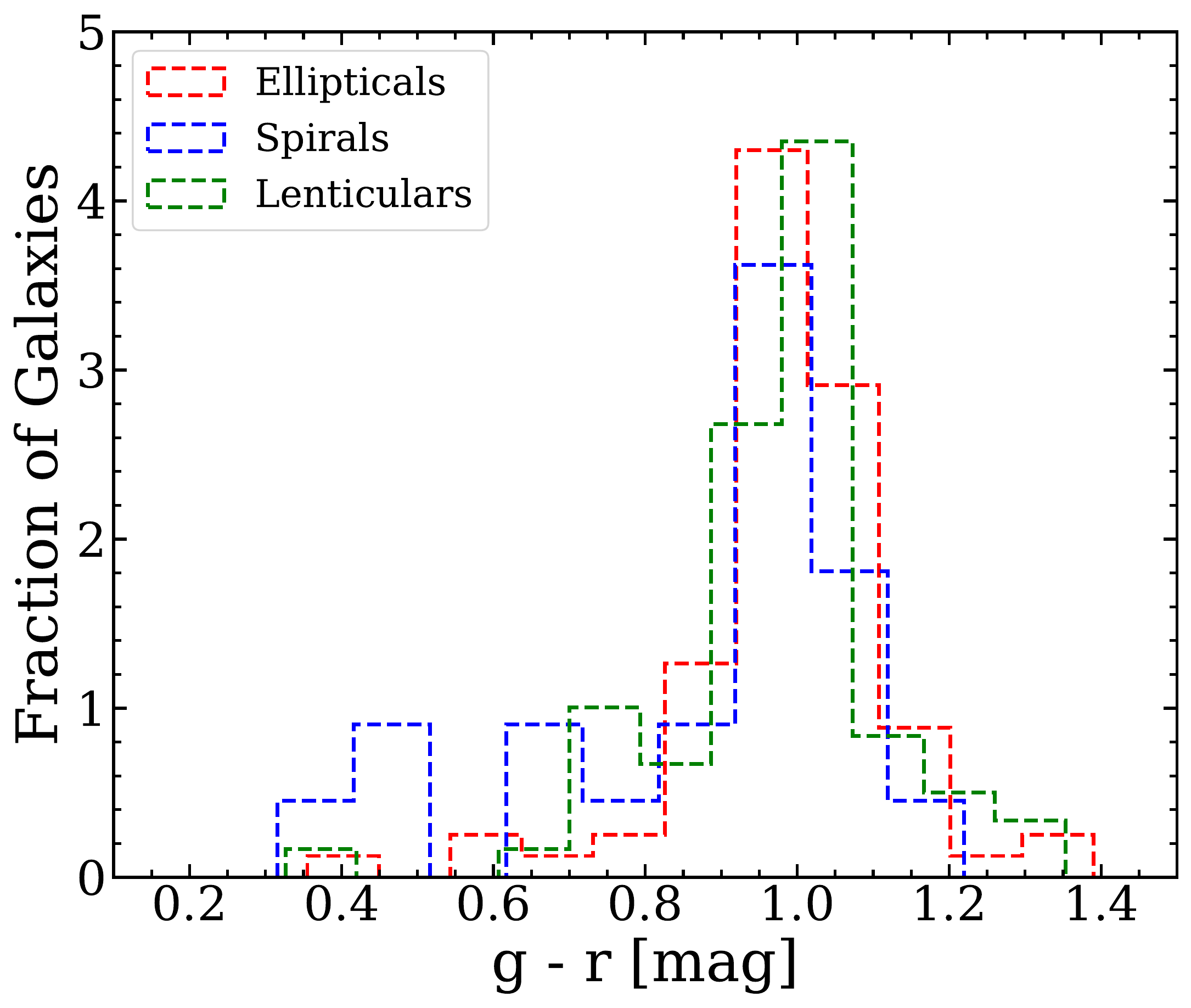}
\includegraphics[width=0.42\textwidth]{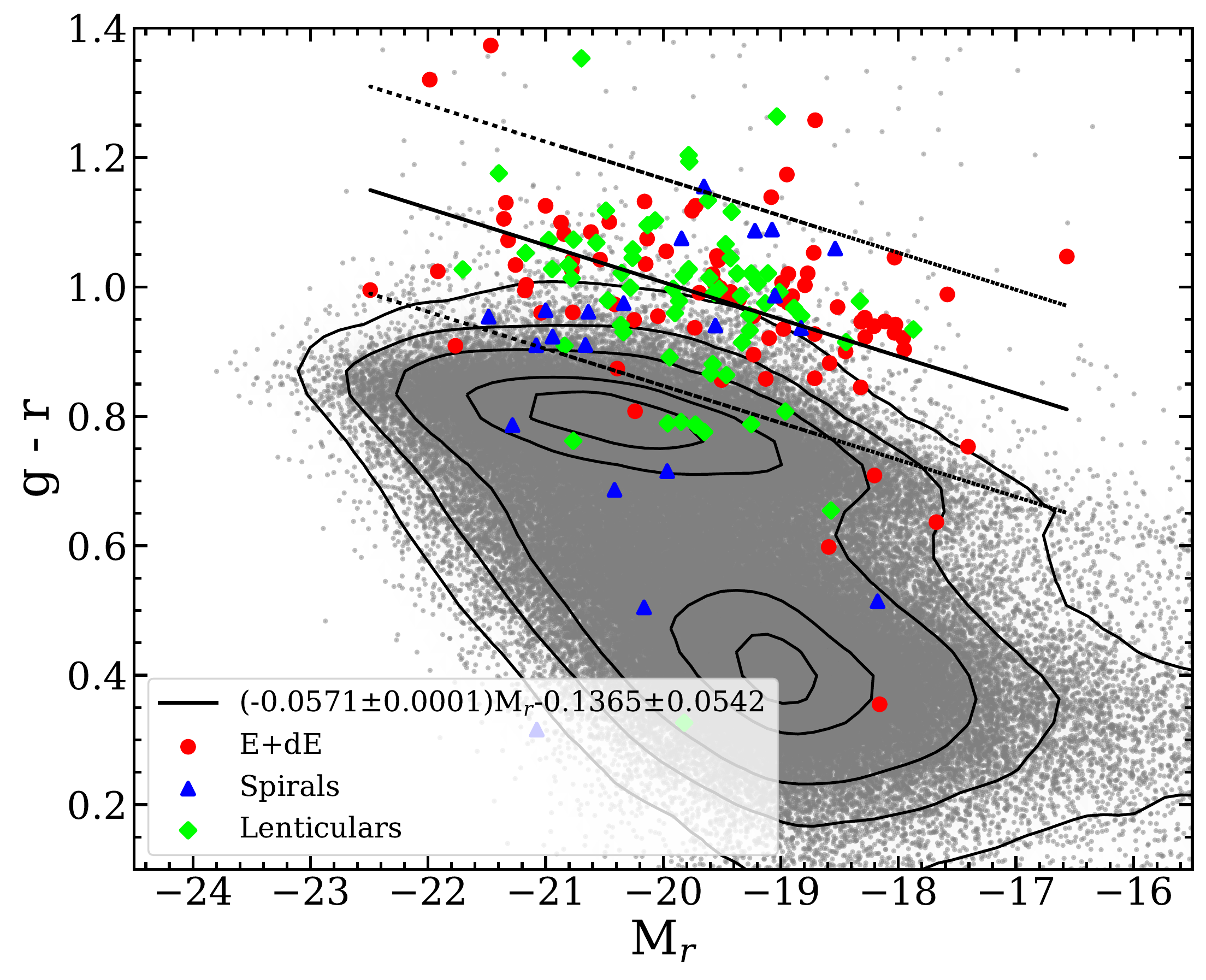}
\caption{ The (Normalised) g-r color histogram of elliptical, lenticular and spiral galaxies in A426 (left). Color-magnitude diagram for our sample of galaxies. The grey dots in the background are galaxies from SDSS sample of spectroscopic galaxies with redshift z < 0.1 and the black-dashed line is a linear fit to the cluster elliptical and dwarf elliptical galaxies and S0s in our sample  (right). Note that all the quantities except background used in these plots are from GALFIT fitting. }
\label{fig:CMR}
\end{figure*}

A small fraction ($\sim 7\%$) of the ellipticals belongs to the compact ellipticals (cE) category of which M32 is prototype \citep{Bekkietal2001,Chilingarianetal2009}.  All 6 cEs are fitted with a Sersic profile and all have Sersic indices $> 3.5$ and r-band absolute magnitude -18 to -20. But their effective radii (median $\sim 1.1$ kpc) are smaller compared to the bright ellipticals (median $\sim 3.2$ kpc) as well as the dwarf ellipticals (median $\sim 1.8$ kpc). 
These median values seem to  suggest us  that the cEs are somewhat smaller in size compared both to the ellipticals and dEs. Various studies suggest that the cEs are likely to have formed by tidal stripping with massive nearby objects - a process that is common in cluster environment \citep{Ferre-Mateuetal2018,Duetal2019}.     

\subsection{Lenticulars}
\label{sec:lenticulars}
 Our sample consists of 64 lenticular (S0) galaxies which are further sub-classified into three different types - $41$ unbarred S0s, $14$ S0s with bar+lens and $9$ dwarf S0s (dS0) with bar+lens. All together, there are $\sim 35\%$ of S0s in our sample. Our S0 fraction (S0 + S0/a + dS0) is similar to that in Coma cluster ($\sim \,31\%$). Interestingly, the S0 fraction can vary from as low as $\sim 15\%$ in Virgo cluster to $\sim 50\%$ in nearby rich clusters \citep{Dressler1980,Binggeli1985,Calvi2012,MichardAndreon2008}. Out of 64 S0 galaxies, $58$ galaxies are fitted with Sersic + Exponential corresponding to a bulge and disk, 6 galaxies are fitted with Sersic + bar + exponential.
 The bulge Sersic index $n_{bulge}$ lies in the range of 0.81 - 5.8 (see Fig.~\ref{fig:struc_param}). About $94\%$ of these S0s have $n_{bulge} < 4.5$, with median $n_{bulge} = 2.7$ --- indicating they might be classical bulges \citep{Barwayetal2016}. Since $n_{bulge} >2.00$ for $75\%$ of the sample, most of our lenticular galaxies are bulge dominated as might be expected in cluster environment where gas stripping and quenching of star-formation might have transformed spirals with bigger bulges to S0s \citep{Laurikainenetal2010, Cortesietal2011}. For more details about the fitting parameters of various morphological types in our cluster, we refer the reader to online tables. Caution to the readers: note that the formal errors computed by GALFIT account only for statistical uncertainties in the flux, therefore the quoted errors are underestimated.  A more realistic estimate of the errors would come from the use of different PSFs, sigma images etc. In the following, we study Color-magnitude relation for the cluster galaxy sample with a focus on ellipticals and S0s.
 
\section{color-magnitude relation}
\label{sec:CMR}
The ETGs, apart from the dwarf population, follow a well-defined color-magnitude relation in a cluster environment. \citet{Sandage1972} and \citet{VisvanathanSandage1977} were among the first to discuss in detail the color-magnitude relation (hereafter, CMR) for ETGs in Coma and Virgo cluster. There seems to be an universal CMR existing for the ETGs in several nearby clusters \citep{Boweretal1992,Boweretal1998}.  
The left panel of Fig.~\ref{fig:CMR} shows the $g-r$ color distribution for our morphologically classified ellipticals, spirals and lenticulars in the cluster. While the dominant population are red as expected from the generic understanding of cluster galaxy evolution \citep{DeLuciaetal2007}, there exists about a dozen bluer population - both bluer ellipticals and spirals but none of the lenticulars except one in the cluster are found to be bluer with ${g-r}<0.6$. When looked at individual galaxies following their colors, it is found that the bluer ellipticals are actually dwarf ellipticals which are classified by SDSS as star-forming. We have examined the spectra and all of them contain emission lines.   
Rest of the ellipticals and lenticulars follow nearly identical color distribution while spirals have slightly blue-ward wing - indicating the evolved state of these galaxies.  

The right panel of Fig.~\ref{fig:CMR} shows the color-magnitude diagram (CMD) for all the visually classified E+dE, lenticulars and spirals. Apart from a handful of bluer spirals and ellipticals, most of the galaxies are at par with the red sequence. Based on the linear regression analysis applied to the E+dE and S0s, it is found that the following relation describes well the CMR:

\begin{equation}
    g-r = (-0.0571 \pm 0.0001) M_{r} - (0.1365 \pm 0.0542),
\end{equation}

\noindent where $M_{r}$ represents the r-band absolute magnitude, spanning over a range of $\sim 6$ mag. The fitted slope of the CMR , $-0.0571$~mag~mag$^{-1}$, is similar to previous estimates of the slope in other cluster environments \citep{Hoggetal2004,Mahajanetal2010,Rocheetal2010}. The scatter in our CMR relation is 0.16, somewhat larger than found in other studies of nearby clusters \citep{Aguerri2020,Agulli_2017,Agulli_2016,Agulli_2014}. The larger scatter might be due to the inclusion of $\sim 2$~mag fainter galaxies. In fact, \cite{Aguerri2020} found a well defined red sequence for A426 only for the brighter magnitudes, $M_{r} < -19$.
Note that most of the elliptical and lenticular galaxies in our cluster follow the same CMR, see \cite{Terlevichetal2001} for CMR on Coma. The overall picture based on the CMR analysis suggests that the cluster seems to lack bluer galaxy population as expected in a normal cluster environment \citep{ButcherOemler1984}. Even the spirals and other disk galaxies are fairly red with $g-r$ color $\sim 1$. Of the 21 spiral galaxies considered here from a total of 23 morphologically classified spirals (barred and unbarred) in our sample, only 5 are late-type with $g-r < 0.8$. In other words,  $\sim 76\%$ of spirals are red with $g-r >0.8$ (see the fourth row of Fig.~\ref{fig:morph}), supporting the so-called Butcher-Oemler effect \citep{ButcherOemler1984} in galaxy clusters. 

The apparent absence of galaxies in the blue cloud indicates strong impact of environment on the star formation. The ram-pressure stripping of gas and galaxy harassment \citep{Mooreetal1996,Abadietal1999}, strangulation \citep{peng2010}, minor mergers \citep{Skeltonetal2009} have probably caused the suppression of star-formation \citep{Poggiantietal1999} and thereby the migration of the late-type galaxies to the red sequence without drastic morphological transformation. Since the cluster has substantial hot gas as seen from the deep X-ray observation by Chandra \citep{Fabianetal2011}, it might also hamper the cooling of the gas and thereby aids in shutting down the star-formation. All these physical processes might have effectively erased the blue cloud population making the cluster galaxy CMD from bimodal to nearly unimodal. Even when the cluster may be in a relaxed state based on the galaxy morphology fraction and CMD evolution \citep{DeLuciaetal2007}, the deep Chandra observation indicates disturbed central region due to the active galactic nucleus of NGC 1275 \citep{Fabianetal2011}.

\section{Scaling relations for the ETGs}
\label{sec:scaling}
Scaling relations are one of the most revealing empirical relations between the macroscopic physical properties of galaxies such as their luminosity, stellar mass, effective radius, velocity dispersion etc., especially to understand the population of a certain morphology class as a whole. For the ETGs, scaling relations provide a great way to decode the formation and evolution of their stellar population. For example, scaling relations allow us to test whether the old stellar population of ETGs formed in a single burst several Gyrs ago \citep{TinsleyGunn1976,Bruzual1983,Trageretal2000,Sanchez_blazquez2006}. Scaling relations are also useful diagnostic to disentangle galaxy properties in high (e.g., cluster) and low (field) environment \citep{deCarvalhoDjorgovski1992,Evstigneevaetal2002,Nigoche-Netroetal2008}. In this section, we study two important scaling relations - Kormendy relation(hereafter, KR) \citep{Kormendy1977} and Fundamental Plane relation(hereafter, FPR) \citep{Terlevichetal1981,Djorgovski1987,Dressler1987} for our ETGs. In order to proceed forward, we apply a number of corrections to the derived physical parameters of the ETGs in our sample. 

First, we correct the effective radii obtained from the modelling of the ETGs observed by SDSS telescope for the inclination effect following the relation \citep{Saulderetal2013},

\begin{equation}
    R_{e} = R_{e,obs} \sqrt{b/a},
    \label{eq:Re_corr}
\end{equation}

\noindent where $b/a$ and $R_{e,obs}$ denote respectively the apparent axis ratio and the effective radius of ETGs obtained from the GALFIT modelling.

\begin{figure}
\includegraphics[width=0.5\textwidth]{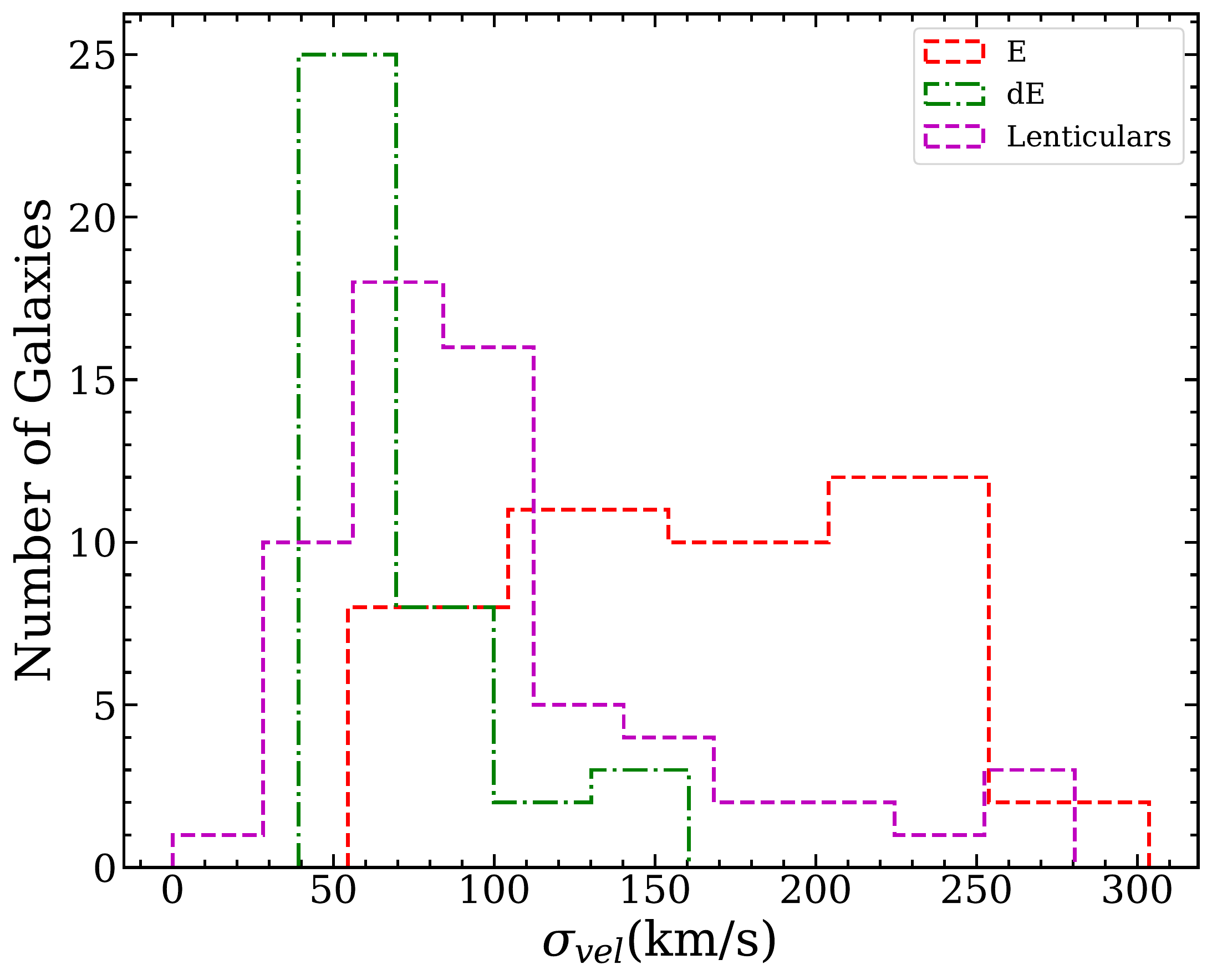}
\caption{The central velocity dispersion histograms of Es, dEs and S0s in our sample. The median values of $\sigma_{vel}$ for Es, dEs and S0s are 164.21~ kms$^{-1}$, 61.68~kms$^{-1}$ and 88.36~kms$^{-1}$ respectively.}
\label{fig:sigma0}
\end{figure}

\begin{figure*}
\includegraphics[width=\textwidth]{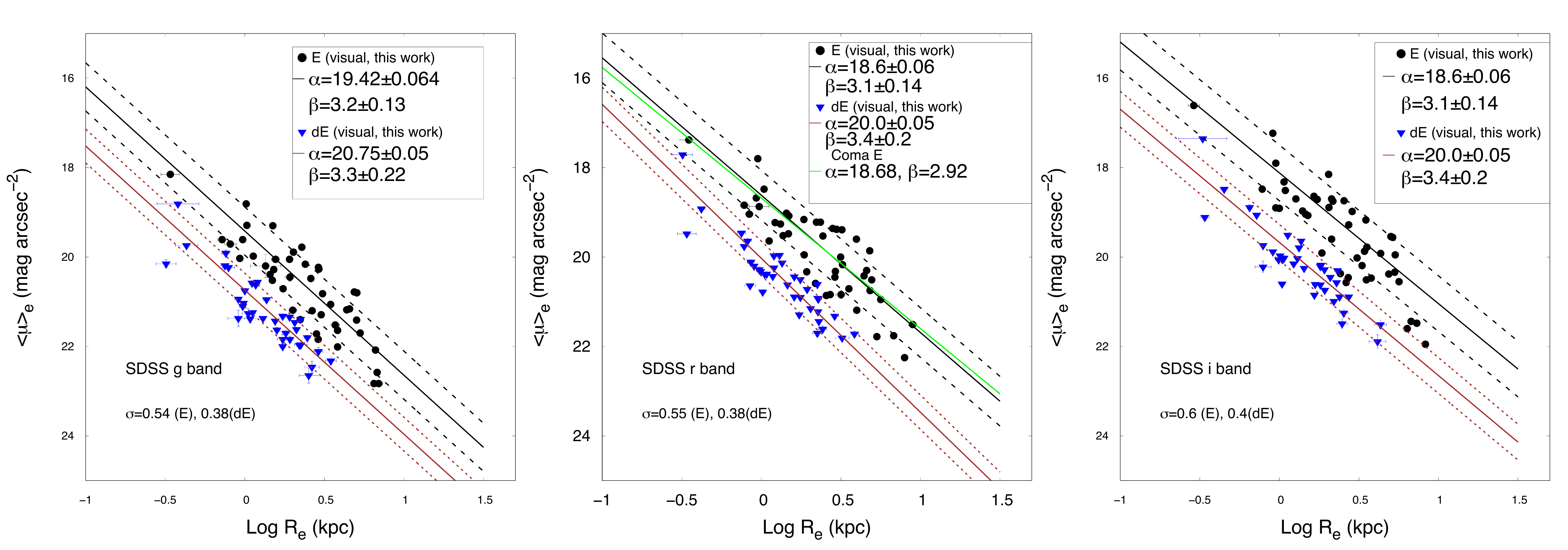}
\caption{Kormendy relation for the Sample1 of visually selected Es and dEs. The inset in each panel represents the fitted Kormendy parameters. The dashed lines represent $1\sigma$ (=root mean square error) scatter from the fitted model and the value of $\sigma$ is mentioned in each panel. Kormendy relation for the Coma cluster ellipticals \citep{LaBarberaetal2003} is shown (by green line) for the sake of comparison only.} 
\label{fig:kormendy_visual}
\end{figure*}

\par
\noindent The mean effective surface brightness ${\langle\mu\rangle}_{e}$ is related to the integrated magnitude and effective radius $R_{e}$ of the galaxy \citep{Graham2005}. Although the integrated magnitudes $m$ for our ETGs are already foreground extinction \citep{Schlegeletal1998} and K-corrected \citep{Pence1976,Jorgensenetal1992,BruzualCharlotte2003}, they might still be subject to changes due to the stellar evolution correction \citep{Poggianti1997} over the look-back time corresponding to the redshift of the cluster. Since our cluster is at a lower redshift ($z \sim 0.016$) and $\Delta z < 0.04$ for all the member galaxies of the cluster, the evolutionary correction term could be neglected without significantly affecting the final outcome. Finally, the mean effective surface brightness used in the following scaling relations is given by

\begin{equation}
    {\langle\mu\rangle}_{e} = m + 2.5log(2\pi{R}_{e}^{2}) -10 \log(1+z), 
    \label{eq:mean_mue}
\end{equation}

\noindent where the last term is due to the cosmological surface brightness dimming \citep{Tolman1930,Tolman1934, HubbleTolman1935,Jorgensenetal1995} in that the surface brightness decreases by a factor of $1/(1+z)^4$, z being the redshift of the galaxy.

The SDSS central velocity dispersion is measured within a fixed aperture of 3" diameter centered on each galaxy. Since the 3" fibre covers different physical area in galaxies at different distances, the observed velocity dispersion needs an aperture correction \citep{Bernardi2007}. We correct the observed central velocity dispersion for our ETGs using the following relation \citep{Jorgensen1996,Bernardi2003a, Bernardi2007, Saulderetal2013, Hou2015} 

\begin{equation}
   \sigma_{vel} = \sigma_{obs}\left(\frac{8 R_{fiber}}{R_e}\right)^{0.04},
   \label{eq:sigma_corr}
\end{equation}

\noindent where $\sigma_{obs}$ is the observed velocity dispersion within the SDSS fibre. $R_{fiber}$ = 1.5" is the angular radius of the fiber and $R_e$ is the effective radius in arcsec. For further details on the SDSS velocity dispersion correction, the readers are referred to \cite{Bernardi2003a}. 

\begin{figure*}
\includegraphics[width=0.9\textwidth]{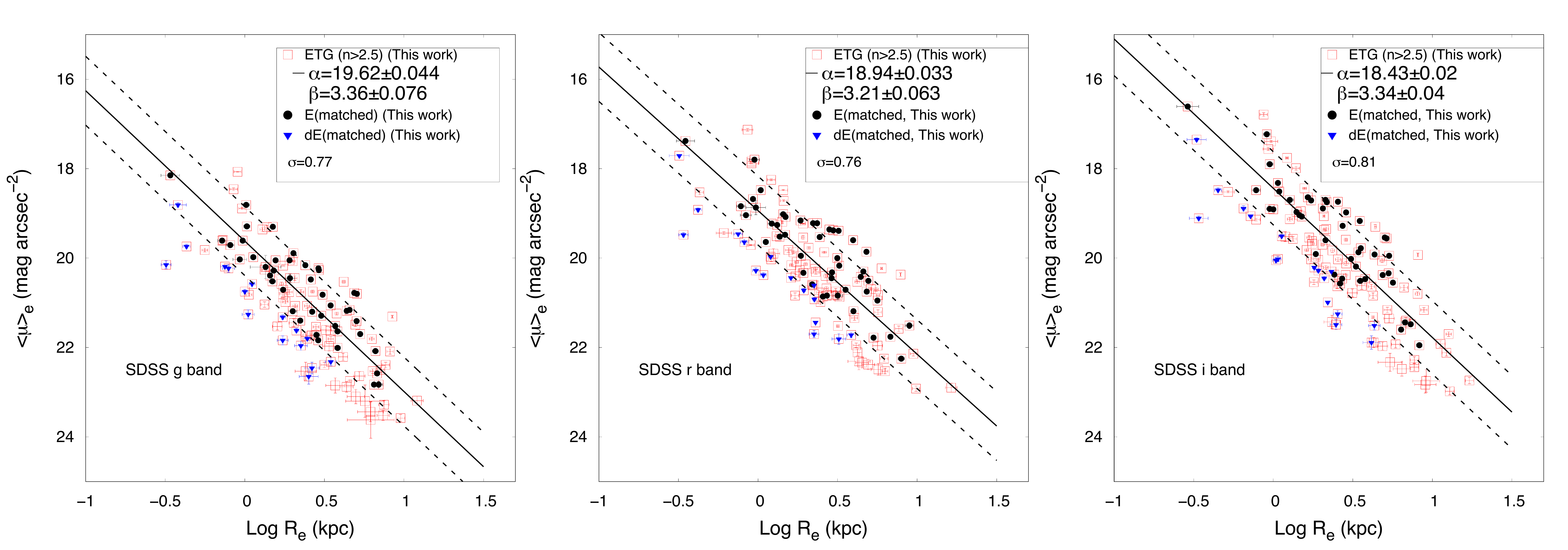}
\includegraphics[width=0.9\textwidth]{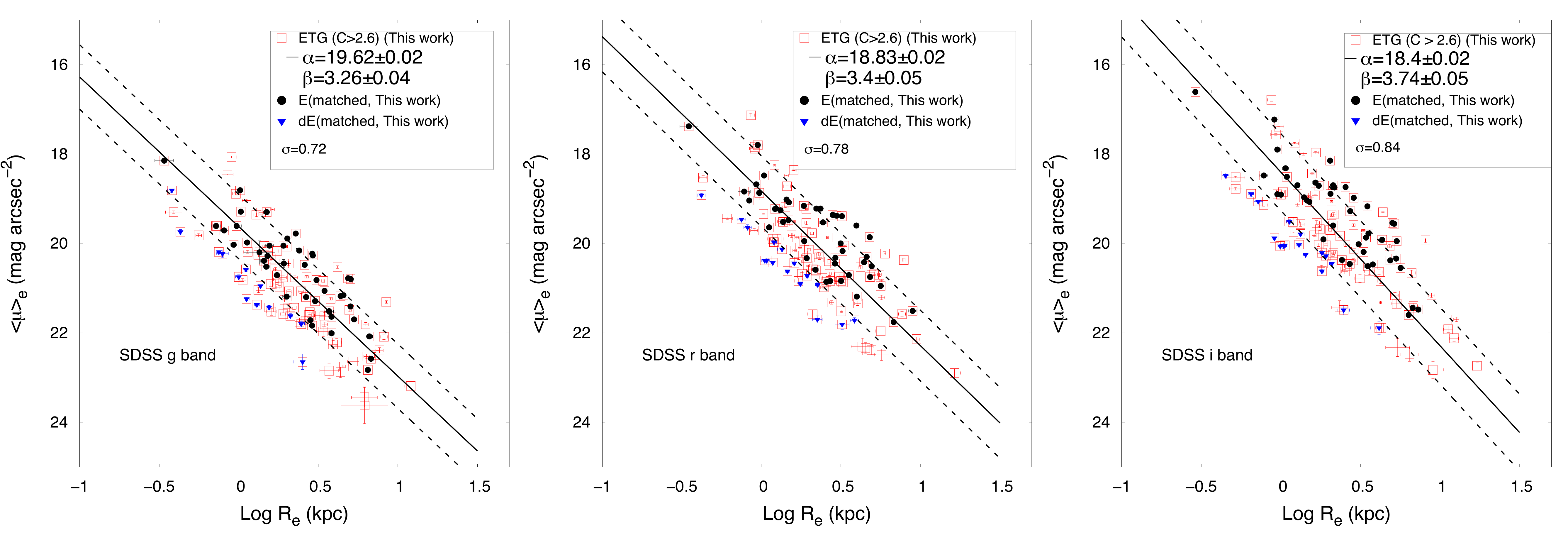}
\includegraphics[width=0.9\textwidth]{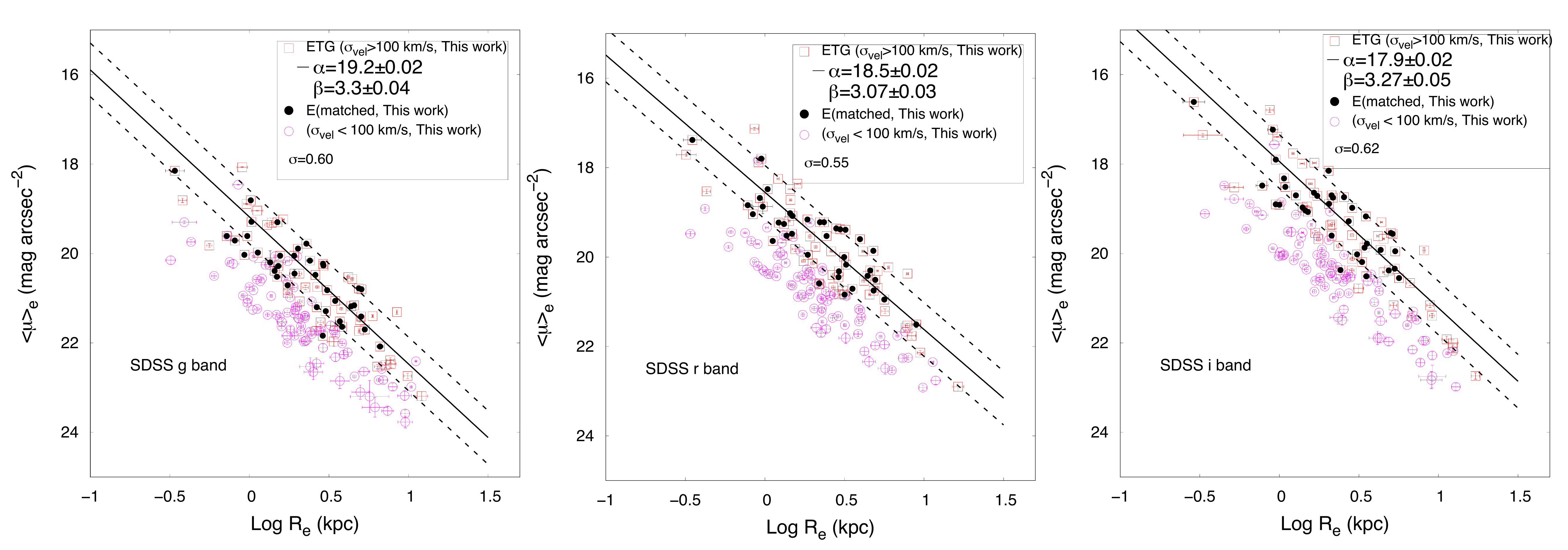}
\caption{Same as in Fig.~\ref{fig:kormendy_visual} but for ETGs in Sample2 (upper panel), Sample3( middle panel) and Sample4 (bottom panel). The scatter, $\sigma$ for each pass band is mentioned in their respective panel.}
\label{fig:kormendy_nCvel}
\end{figure*}

\subsection{Construction of ETG Samples}
\label{sec: samples}
There are several ways to construct ETG samples which can be used to probe galaxy scaling relations. For local galaxies, it is possible to examine the morphology visually. Although, it is non-trivial to have an unbiased sample selection in the higher redshift where images are resolution limited. However, learning from the low-redshift universe would be helpful in dealing with the high-z galaxies. For our spectroscopically confirmed cluster members, we define samples in the following five different ways:\\

\noindent \textbf{Sample1:} This is a sample of Ellipticals based on visual morphology guided by visual inspection of galaxy fitting residuals. All these galaxies were fitted well with single Sersic component. Of the 84 galaxies in this sample, 43 are normal Es and 41 dEs. Only 38 dEs are used in KR and FPR as 3 of them have zero measured velocity dispersion. \\

 \noindent \textbf{Sample2:} This sample of ETGs consist of galaxies with Sersic index, $n \geq 2.5$ \citep{Blanton_2003,Tortorellietal2018, mcintosh_2005}. Since Sersic index varies with wavelength, there are  127, 129 and 132 galaxies in this sample in g, r and i-bands respectively. All fittings are based on single Sersic model. In r-band, these include 43 Es and 16 dEs from Sample1. All the remaining dEs have $n < 2.5$ and does not form part of this sample. As shown by \cite{Tamburri_2014}, it is possible that real morphologically classified elliptical galaxies can have $n < 2.5$. Remaining 68 galaxies include S0s, dSphs.\\ 
 
 \noindent \textbf{Sample3:} This sample of ETGs is based on concentration index, $C_{95} \equiv R_{p90}/R_{p50}$. Following \cite{Stratevaetal2001}, we define ETGs for which $ C_{95} > 2.6$. There are 125, 128 and 129 galaxies in g, r, i-bands respectively which satisfy this criterion.  In r-band, there are 42 Es and 16 dEs common with Sample1. \\
 
\noindent \textbf{Sample4:} Here, we divide the morphologically selected ETGs into two categories - one with $\sigma_{vel} > 100$~km~s$^{-1}$ \citep{Gargiulo_2009} and other with $\sigma_{vel} < 100$~km~s$^{-1}$. We have chosen 100~km~s$^{-1}$ as the separator between low-$\sigma_{vel}$ and high-$\sigma_{vel}$ galaxies based on the dE central velocity distribution, see Fig.~\ref{fig:sigma0}. Most of the dEs (33) have $\sigma_{vel} < 100$~km~s$^{-1}$ and only 5 dEs have $\sigma_{vel} > 100$~km~s$^{-1}$. With this value of the velocity dispersion as the separator, 7 Es and 36 S0s have $\sigma_{vel} < 100$~km~s$^{-1}$ and 36 Es and 26 S0s have $\sigma_{vel} > 100$~km~s$^{-1}$ in r-band. There are 68, 67 and 68 ETGs in this sample in g, r and i-bands respectively in the first category. In the r-band, this sample include 37 Es and 5 dEs.\\ 

\noindent 
\textbf{Sample5:} Here, we select only those galaxies for which $n \geq 2.5$, $C_{95} > 2.6$ and $\sigma_{vel} > 100$~km~s$^{-1}$ are satisfied simultaneously. We do this for r band only. In the r-band, 70 galaxies satisfy this criterion, which include among others 37 Es and 4 dEs. 7 E galaxies and 34 dE galaxies fail to satisfy at least one condition. See Table~\ref{tab: kormendy_parameters} for details.

\noindent 
Complimentary Sample5: Galaxies satisfying $n < 2.5$, $C_{95} < 2.6$ and $\sigma_{vel} < 100$~km~s$^{-1}$ condition  simultaneously. 
\\

\subsection{Kormendy Relation}
\label{sec:kormendy}

Luminous ellipticals are larger in size and have lower effective surface brightness compared to their low luminous counterpart - indicating a relation between their effective radius $R_{e}$ and the mean surface brightness ${\langle\mu\rangle}_{e}$ within that radius, known as the KR \citep{Kormendy1977}. The ETGs, classical bulges of spiral and lenticular galaxies, in the local universe as well as at intermediate redshifts, are known to follow this correlation tightly \citep{Capacciolietal1992,LaBarberaetal2003,Redaetal2004,Longhettietal2007,Tortorellietal2018}. Since it relates the distribution of light and size of an ETG, the correlation has also been used extensively to understand the evolution of the ETGs and in particular, the size evolution over the cosmic time \citep{Trujillo2006, Longhettietal2007, Rettura2010, Trujillo2011, NaabOstriker2017}. Nevertheless, several things remain unclear, e.g., how the Kormendy relation varies across different environments in the local universe as well as in the intermediate redshift range; whether there is a cluster to cluster bias; how it depends on the evolutionary stages of the ETGs, colors, magnitude range, wavelength \citep{Bernardietal2003b,Nigoche-Netroetal2007,LaBarberaetal2010,Samiretal2020}. We study here the KR for five different samples. We fit the following form of the KR to our samples of ETGs \citep{Kormendy1977}: 
 
 \begin{equation}
     {\langle\mu\rangle}_{e} = \alpha +  \beta \log{R_{e}},
     \label{eq:eqkormendy}
 \end{equation}
 
    \noindent where ${\langle\mu\rangle}_{e}$ is the mean effective surface brightness computed within $R_e$ and corrected for the cosmological surface brightness dimming (see Eq.~\ref{eq:mean_mue}). The slope $\beta \simeq 3$ as shown by \citep{Kormendy1977}. Both the slope and the intercept might vary depending on the passband used for the observation of the galaxies. Fig.~\ref{fig:kormendy_visual} shows the KR for our {\it Sample1} consisting of normal Es and dEs. Based on the fitting that uses the Levenberg-Markwardt non-linear least square method \citep{Markwardt2009}, we obtain a slope of $\beta = 3.1\pm0.14$ and an intercept of $\alpha=18.62\pm0.06$ for the SDSS r band. The slopes for normal Es is consistent within errors as a function of wavelength. The KR slope for Es matches within errors to that of Coma Es in the r-band \citep{LaBarberaetal2003}. The intercept (zero-point) varies from $19.42$ to $18.60$ as function of wavelength (from g- to i- band), possibly due to the stellar population, but the exact reason for this apparently systematic variation remains to be investigated. (see Table~\ref{tab: kormendy_parameters} for details).

To study whether the sample selection impacts our KR estimate, we compare the KRs built with all the samples defined in section~\ref{sec: samples}. Fig.~\ref{fig:kormendy_nCvel} shows the KR for Sample2 (based on $n>2.5$), Sample3 (based on $C>2.6$) and Sample4 (based on  $\sigma_{vel}>100$~kms$^{-1}$). The fitted slope and intercept along with other parameters are presented in Table~\ref{tab: kormendy_parameters}. It appears that the scatter $\sigma$ (=root mean square error, RMSE) is lowest ($=0.55$) for Sample4. In other words, the sample of ETGs selected based on visual + kinematics is preferred over the Sersic index or concentration index alone. The reason for this is that adding an additional constrain reduces the scatter and because the additional constraint is the third direction of the FPR. 
As shown in Fig.~\ref{fig:Chist_new} and Fig.~\ref{fig:struc_param}, Sample2 and Sample3 contains a large number of S0 galaxies which are disk dominated, inclusion of which would result in increasing the net scatter. Of the 70 ETGs in Sample4, there are 37 Es (common from Sample1). These 37 Es show the least scatter ($\sigma =0.47$) in the KR; the best fit slope and intercept are $\beta =2.91 \pm 0.07$ and $\alpha = 18.64$ which are close to the best-fit parameters of the full sample4. Both the slope and intercept of our cluster Es agree well with the Coma cluster ellipticals ($\beta_{Coma} =2.92$, see \cite{LaBarberaetal2003}), especially for Es selected under Sample4 in r-band.
The slopes and scatter of the KR in r-band for our sample1 and sample4 are consistent with the ETG, red and passive sample of galaxies belonging to the Frontier cluster AS1063 at z=0.35. The same holds true for the ETG, red and passive sample of M1149 (z=0.55) against our sample1 and sample4 in g-band, see \cite{Tortorellietal2018}. However, the zero-points in our samples differ significantly; they are fainter by $\sim 0.6$~mag in case of AS1063 and by $\sim 2$~mag for M1149. They differ because of luminosity evolution with redshift.   Based on a sample of four clusters at z $\sim$ 0, 0.21, 0.31, 0.64 with a total of N = 228 spheroidal galaxies, \cite{LaBarberaetal2003} found that the slope of the KR remained nearly unchanged (variation from $2.74 - 3.04$ with a typical uncertainty of 0.2) since $z \sim 0.64$ i.e., over the past 6 Gyr. At even higher redshits ($1.2 <z < 1.4$), based on 56 cluster ellipticals, \cite{Saraccoetal2017} found the Kormendy slope to be $\beta$ = 3.0 $\pm$ 0.2. If we consider all five samples of ETGs (presented in Table~\ref{tab: kormendy_parameters}) and the KR for the Ellipticals only, we have the mean slope $\beta = 3.02 \pm 0.1$. A corollary that could be derived from this is that the ETGs in this cluster settled into Kormendy plane long ago and not much evolution has happened at least during the last 6 Gyrs in our sample of ellipticals. This, in turn, could produce powerful constraint on the overall formation of the cluster itself \citep{KravtsovBorgani2012}. 

\par
Dwarf E galaxies are fainter than normal E galaxies \citep{Nigoche-Netroetal2008}. Our cluster dE galaxies are about $2$~mag fainter than their brighter counterpart and less centrally concentrated, having a median Sersic index of n=2.5 as compared to Es with n=4.1 (see Fig.~\ref{fig:struc_param}).  Most of the dEs (from Sample1) have $\sigma_{vel} < 100$~kms$^{-1}$. It appears that the dEs are probably the fainter version of the normal Es and might share similar formation history although the debate is still on \citep{Bosellietal2008}. KR for the dE galaxies shows a clear separation with fainter zero-point (see Fig.~\ref{fig:kormendy_visual}). The dEs in Sample2 and Sample3 are also outside 1$\sigma$ scatter. The bottom panel of Fig.~\ref{fig:kormendy_nCvel} show galaxies that are outside the 1$\sigma$ scatter, are mostly dEs plus a few S0s. They seem to form a parallel KR.

\begin{figure}
\includegraphics[width=0.5\textwidth]{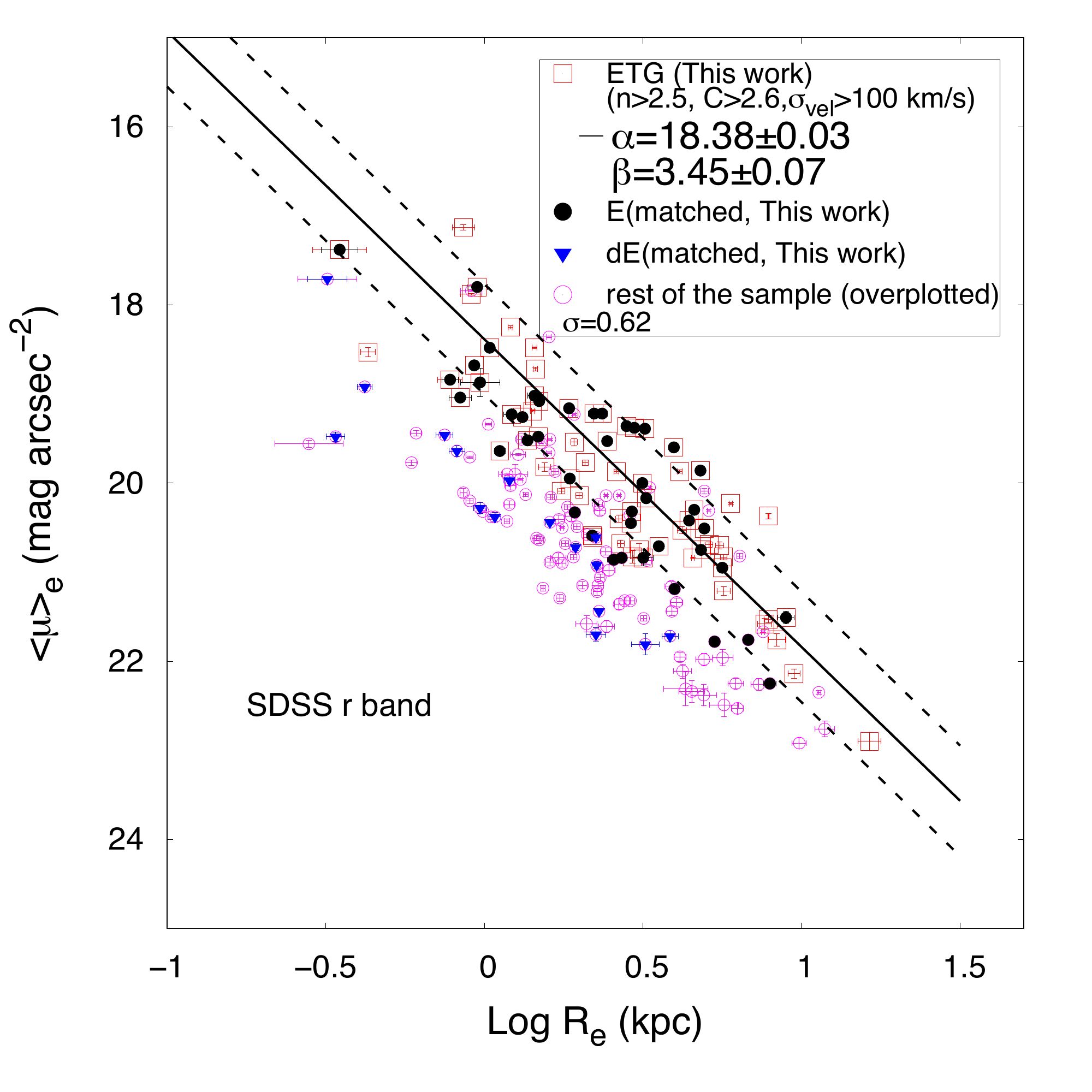}
\caption{The Kormendy Relation for the sample5 of ETGs in the SDSS r band. Over plotted are galaxies from the complimentary sample5. The Es and dEs are from our visually selected sample which are matched. The $1\sigma$ scatter from the fitted KR relation is 0.62.}
\label{fig:KR_sample5}
\end{figure}

Fig.~\ref{fig:KR_sample5} shows the KR for the sample5 in the r-band. Although the slope of the KR for sample5 ($\beta =3.45\pm0.07$) is steeper than that of Sample4 ($\beta =3.07\pm0.03$), the complimentary sample i.e., mostly the low-velocity dispersion galaxies is similar in either case, and they seem to be suggesting a parallel KR.

\begin{table*}
\caption{Kormendy relation(KR) best-fitting  parameters for samples defined in section~\ref{sec:scaling}. Here column 1 shows the various samples used to build KR. Columns 2, 3, 4, 5, $\,$  6, 7, 8, 9 and 10, 11, 12, 13  represent respectively number of galaxies, zero-points, slopes and scatters in g, r and i-bands. Note that error on $\sigma$ in all these cases is $\pm0.01$.}

\begin{tabular}{lcccccccccccl}
\hline
\hline
{}&{}&{}&$g\, band$&{}&{}&{}& $r\, band$&{}&{}&{}&$i\, band${}\\
\hline

$Sample$&{$N$} & $\alpha$ & $\beta$ & $\sigma$ &$N$& $\alpha$ & $\beta$ & $\sigma$ &$N$& $\alpha$ & $\beta$ & $\sigma$ \\
{(1)} & {(2)}& {(3)}& {(4)} & {(5)} & {(6)} & {(7)} & {(8)} & {(9)} &{(10)} & {(11)} & {(12)} & {(13)}\\
\hline
Sample1(E) & 43 & 19.42$\pm$0.06 & 3.22$\pm$0.13 & 0.54 & 43 & 18.62$\pm$0.06 & 3.10$\pm$0.14 & 0.56 & 43 & 18.60$\pm$0.06 & 3.10$\pm$0.14 & 0.62 \\ 
Sample1(dE) & 38 & 20.75$\pm$0.05 & 3.30$\pm$0.22 & 0.38 & 38 & 20.03$\pm$0.05 & 3.44$\pm$0.21 & 0.38 & 38 & 20.00$\pm$0.05 & 3.40$\pm$0.20 & 0.40 \\ 
\hline
Sample2 & 127 & 19.62$\pm$0.04 & 3.36$\pm$0.08 & 0.77 & 129 & 18.94$\pm$0.03 & 3.21$\pm$0.06 & 0.76 & 132 & 18.43$\pm$0.02 & 3.34$\pm$0.04 & 0.81 \\ 
E(matched) & 43 & 19.34$\pm$0.06 & 3.48$\pm$0.12 & 0.54 & 44 & 18.62$\pm$0.05 & 3.22$\pm$0.10 & 0.55 & 43 & 18.21$\pm$0.06 & 3.41$\pm$0.11 & 0.56 \\ 
dE(matched) & 16 & 20.66$\pm$0.10 & 3.13$\pm$0.31 & 0.42 & 16 & 19.94$\pm$0.07 & 3.28$\pm$0.19 & 0.41 & 17 & 19.53$\pm$0.07 & 3.21$\pm$0.21 & 0.43 \\ 
\hline
Sample3 & 125 & 19.62$\pm$0.02 & 3.26$\pm$0.04 & 0.72 & 128 & 18.83$\pm$0.02 & 3.40$\pm$0.05 & 0.78 & 129 & 18.40$\pm$0.02 & 3.74$\pm$0.05 & 0.84 \\ 
E(matched) & 42 & 19.37$\pm$0.05 & 3.34$\pm$0.11 & 0.53 & 42 & 18.72$\pm$0.04 & 3.04$\pm$0.09 & 0.52 & 42 & 18.11$\pm$0.05 & 3.29$\pm$0.10 & 0.56 \\ 
dE(matched) & 13 & 20.67$\pm$0.08 & 3.70$\pm$0.38 & 0.28 & 16 & 20.02$\pm$0.08 & 3.08$\pm$0.32 & 0.25 & 17 & 19.67$\pm$0.07 & 3.16$\pm$0.34 & 0.28 \\ 
\hline
Sample4 & 68 & 19.20$\pm$0.02 & 3.30$\pm$0.04 & 0.6 & 67 & 18.50$\pm$0.02 & 3.07$\pm$0.03 & 0.55 & 68 & 17.90$\pm$0.02 & 3.27$\pm$0.05 & 0.62 \\ 
E(matched) & 37 & 19.22$\pm$0.05 & 3.46$\pm$0.11 & 0.53 & 36 & 18.64$\pm$0.03 & 2.91$\pm$0.07 & 0.47 & 36 & 18.14$\pm$0.04 & 2.76$\pm$0.08 & 0.49 \\ 
\hline
Sample5 & {} &  &  & {} & 70 & 18.38+0.03 & 3.45+0.07 & 0.62 & {} &  &  & {} \\ 
E(matched) & {} & {} & {} & {} & 37 & 18.58+0.03 & 2.85+0.07 & 0.48 &{} & {} & {} & {} \\ 
\hline
\end{tabular}
\label{tab: kormendy_parameters}
\end{table*}

\begin{figure*}
\includegraphics[width=1.05\textwidth]{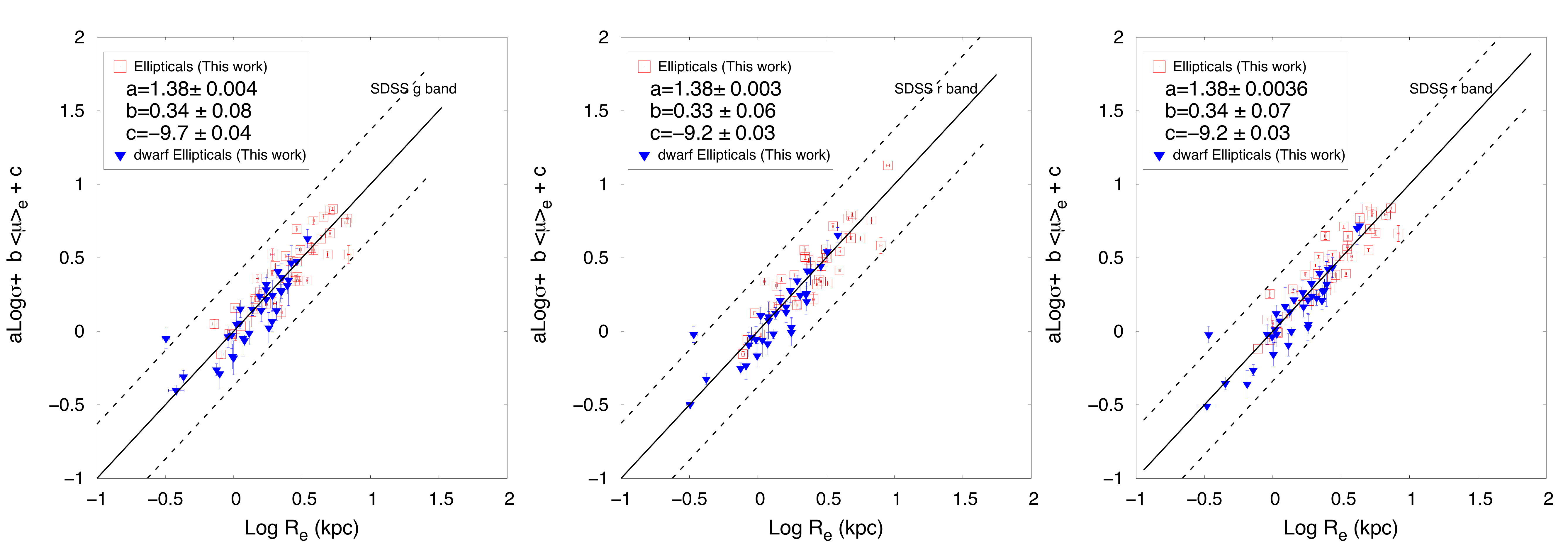}
\caption{Edge-on projection of the Fundamental Plane relation for the visually selected sample of ellipticals and dwarf ellpticals (same as in sample1 in Fig.~\ref{fig:kormendy_visual}) in the SDSS gri bands . The corrected central Velocity dispersions are in km/s, and mean  surface brightness is in $mag/arcsec^{2}$. The $1\sigma$ scatter from the fitted relation is mentioned in Table~\ref{tab:FP} for each band. }
\label{fig:FP_sample1}
\end{figure*}

\subsection{Fundamental Plane Relation}
\label{sec:FP}

\begin{figure}
\includegraphics[width=0.5\textwidth]{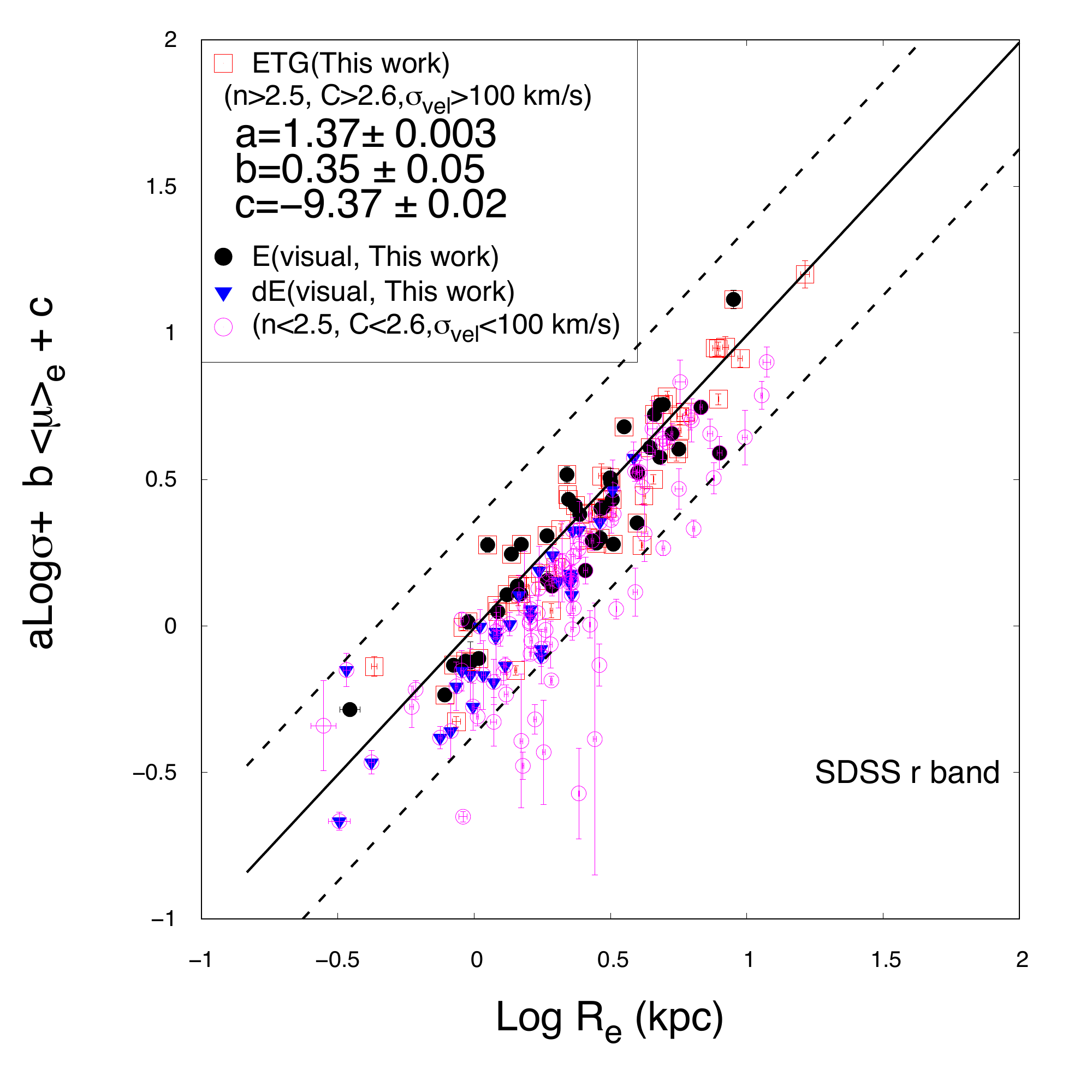}
\caption{The Fundamental Plane relation for the sample5 of ETGs in the SDSS r-band. Over plotted are the galaxies from complimentary sample5. The Ellipticals and dwarf ellipticals are from our visually selected sample which are matched. The $1\sigma$ scatter from the fitted FP relation is 0.36.}
\label{fig:FP_sample5}
\end{figure}

It is well known that the ETGs follow one of the tightest correlations between their structural parameters derived from modelling the surface brightness distribution and the kinematics of their stars - known as FPR \citep{Terlevichetal1981,Djorgovski1987,Dressler1987,Luceyetal1991,Robertsonetal2006}. The FPR for the ETGs is defined by the effective radius $R_e$, the mean effective surface brightness ${\langle \mu \rangle}_{e}$ (corrected for the surface brightness dimming) and central velocity dispersion, $\sigma_{0}$ of their stars as

\begin{equation}
    \log(R_{e}) = a \log{\sigma_{0}} + b {\langle{\mu}\rangle}_{e} + c,
    \label{eq:FPR}
\end{equation}

\noindent where $a$, $b$ and $c$ are constants to be determined for a given sample of ETGs. As shown by Dressler and Djorgovski \citep{Dressler1987,Djorgovski1987}, the FPR has very little scatter compared to other scaling relations suggesting their unique role in unfolding the formation scenario of the ETGs. However, a number of subsequent studies have looked into the FPR and found non-negligible scatter in the relation \citep{Jorgensenetal1993}, although the overall picture of the tightness of FPR compared to other scaling relations remains intact. If the ETGs are strictly following virial theorem for their structural evolution, the values of $a = 2$ and $b = 0.4$. However, in observations, the ETGs follow an FPR with different $a=1.24$ and $b=0.328$ \citep{Jorgensen1996,Busarello1997}. The difference between the observed FPR and the one that is predicted by the virial theorem is known as the tilt of the Fundamental Plane. The age of the stellar populations, mass-to-light ratio (M/L), metallicity, shape of the light profiles could cause the observed tilt of the FPR \citep{Prugniel1994,Graham1997,AguerriGonzalez-Garcia2009}. 
Fig.~\ref{fig:FP_sample1} shows the FPR for the normal and dwarf ellipticals as in Sample1 in the SDSS g,r,i-bands. The fitted parameters for the normal ellipticals ($a$ and $b$) vary little across bands, but the zero point $c$ does instead, see Table~\ref{tab:FP}). Although, the dEs seem to follow the same FPR as followed by the normal ellipticals, we notice a tendency to deviate on the low-mass side since most dEs in our sample have $\sigma_{vel}<100$~kms$^{-1}$. Our derived relation might support a plausible structural transformation between the bright and dwarf ellipticals in the hostile cluster environment \citep{AguerriGonzalez-Garcia2009}. It is likely that the bright ellipticals have faded to join some of these dwarf E population in this cluster although an alternate scenario in which dwarfs form through a different channel might exist. In either scenario, we would require chemodynamical modelling using Integral field spectroscopy to firmly establish that.
Finally, we pick Sample5 which simultaneously satisfy $n>2.5$, $C>2.6$ and $\sigma_{vel}>100$~kms$^{-1}$ providing us with 70 ETGs of which 37 are our visually selected Es. By construction, most of dEs are missed out here. Fig.~\ref{fig:FP_sample5} shows the FPR for this sample of ETGs. The FPR parameters of Sample5 are in close agreement with that of Sample1 of the visually selected Es. The complimentary sample, especially the low-mass systems \citep{Gargiulo_2009} including dEs, dSphs, dS0s, although they are early type galaxies strongly deviate from the FPR of the Sample5. Note that departure from a linear trend of the FPR has been indicated previously by several authors e.g., \cite{Desrochesetal2007,DOnofrioetal2008,Nigoche-Netroetal2008, Hyde2009} found that the inclusion of faint galaxies influences one of the coefficient of the FPR. However, much work is left to be done to arrive at a conclusive remark \citep[see][]{Hyde2009}.

\begin{table*}
\begin{minipage}{175mm}
\caption{Fundamental plane parameters of the samples . Here column 1 shows the samples used to build FP. Columns 2, 3, 4, 5, $\,$  6, 7, 8, 9 and 10, 11, 12, 13  represent respectively slopes corresponding to velocity dispersion, slopes corresponding to mean surface brightness, intercepts and scatters in g, r and i-bands. Note that error on $\sigma$ in all these cases is $\pm0.001$. }
\scriptsize
\begin{tabular}{@{}lccccccccccccc@{}}
\hline
\hline
{}&{}&$g\, band$&{}&{}&{}& $r\, band$&{}&{}&{}&$i\, band${}\\
\hline

$Sample$ & $a$ & $b$ & $c$ & $\sigma$ & $a$ & $b$ & $c$ & $\sigma$ & $a$ & $b$ & $c$ & $\sigma$ \\
{(1)} & {(2)}& {(3)}& {(4)} & {(5)} & {(6)} & {(7)} & {(8)} & {(9)} &{(10)} & {(11)} & {(12)} & {(13)}\\
\hline
Sample1(E) & 1.38$\pm$0.004 & 0.34$\pm$0.08 & -9.68$\pm$0.04 & 0.123 & 1.38$\pm$0.003 & 0.33$\pm$0.06 & -9.18$\pm$0.03 & 0.125 & 1.38$\pm$0.004 & 0.34$\pm$0.07 & -9.18$\pm$0.03 & 0.114 \\ 
Sample5 & {}& {}&{}& {} & 1.37$\pm$0.003 & 0.35$\pm$0.05 & -9.37$\pm$0.02&{}&{}&{}&{}&{}\\
\hline
\end{tabular}
\label{tab:FP}
\end{minipage}
\normalsize
\end{table*}

\section{Discussion and Conclusions}
\label{sec:concl}
    We have studied in detail the morphology of 183 member galaxies in A426 having spectroscopic redshift measurements from the SDSS survey. Followed by the visual morphology, we define 5 different samples of ETGs based on visual morphology, Sersic index, concentration parameter and central velocity dispersion and compare their scaling relations.
Although, there are many more photometric member galaxies belonging to this cluster, we focused on this sample of 183 galaxies, as our primary aim has been to investigate the detailed morphology, scaling relations requiring velocity dispersion measurements. Our analyses show that more than 80\% of total galaxies are early type i.e., Ellipticals (49\%) and lenticulars (35\%) suggesting evolved nature of the cluster galaxy population. Most of the spiral galaxies in the cluster are red with ${g-r} >0.8$, as expected in a cluster environment \citep{ButcherOemler1984}. 

GALFIT \citep{Peng2002,peng2010} is an excellent tool to perform 2D modelling of a galaxy's surface brightness distribution. However, a lot of care must be taken in using it properly. We have realized that cutout size of the galaxy stamps if less than $3\times$~Petrosian radius, leads to overestimation of the fitting parameters. The effective radii and the Sersic indices are more severely affected than the integrated magnitudes. While performing GALFIT for cluster galaxies, it is extremely important to take into account the intracluster light (ICL) or light from the bright nearby galaxies or saturated stars. We have attempted to include the sky component mimicking the ICL \citep{Fischer2017}. The effect of this is severe in the surface photometry of the dwarf ellipticals in the sample. We have removed 4 (mentioned earlier) galaxies including two peculiar galaxies from the sample while performing the GALFIT modelling. An intriguing issue while doing the GALFIT is the smallness of the quoted error bars. This is because the formal errors computed by GALFIT account only for statistical uncertainties in the flux and thereby underestimating the errors. A more realistic estimate of the errors would come from the use of different PSFs, sigma images, different sky background estimates, choosing the accurate model for the fit etc. \citep{Tortorellietal2018, Peng2002, peng2010} Although GALFIT residuals are useful guide to choose a best-fit, it is unclear if that could be used to measure errors in the fitting parameters. 

A large number of studies based on nearby galaxy clusters suggest that dwarf ellipticals are abundant in cluster environment and are fainter than the bright ellipticals \citep{Ferguson1994, Nigoche-Netroetal2008,hashimotoetal2018}. Previously, several studies aimed at understanding whether dEs are a distinct class of galaxies compared to the normal ellipticals and follows a  different formation scenario, but the debate is still on \citep{Kormendy1985,Graham_2003,Graham_2005,Tolobaetal2012}. 
Most ellipticals in our sample belong to the cluster red sequence. However, the cluster seems to have also a few comparatively bluer ellipticals albeit with higher concentration. We have checked that these bluer ellipticals (about 5 of them) have no preferred Hubble type or apparent axis ratios and these are identified as the dwarf ellipticals. The bluer dEs might indicate recent activity or evidence of the accretion of cold gas \citep{Boehringeretal1993} which might have triggered the star-formation in these galaxies. This will be followed up in a future paper.

The cluster has a significantly low bar fraction $\sim 17\%$ including bars and lenses. However, among the 64 S0s, we have about 23 S0s+dS0s with bars and lenses - this is about $36\%$ of all the S0s in our sample. This fraction is similar to what is found in the Coma cluster \citep{Marinovaetal2012}. Of course, the actual bar fraction is even lower if we remove the lenses from our sample and this might be blamed on the cluster environment effect \citep{Butaetal2010, Barwayetal2011}. It is possible that some of the bars are transformed to lenses through interaction and minor mergers, see \cite{Ghoshetal2021} and references therein. 
 
Scaling relations such as KR and FPR depends on the sample of ETGs. Based on an extensive analysis (previous plus this work) of the KR, it is found that the slope of the KR is nearly universal since about redshift $z\sim 1.3$ \citep{LaBarberaetal2003,Saraccoetal2017,Tortorellietal2018}. 
The slope of the KR for dEs is nearly the same as that of the normal Es but with a fainter intercept or zero-point, in other words, the dEs seem to follow a parallel KR. It remains to be investigated whether this could be translated to their different formation scenario \citep{Bosellietal2008} compared to the bright or massive ellipticals. The dEs also seem to follow the same fundamental plane relation as followed by the bright ellipticals. However, we have noticed that some dEs and other ETGs with low velocity dispersion (pop up in the complimentary sample shown in Fig.~\ref{fig:KR_sample5} and Fig.~\ref{fig:FP_sample5})  tend to deviate from the linear FPR. They are also the ones that form the parallel KR. Our studies indicate possible transformation of the bright Es to dwarf Es in the cluster environment \citep{AguerriGonzalez-Garcia2009} or a different formation channel but future studies based on IFU kinematics and chemodynamical modelling would be essential to understand the actual physical processes.\\

\noindent Conclusions from our analysis of A426 are the following:\\

\noindent {\bf 1.} Our spectroscopic sample of 183 member galaxies in A426 is spanned over a range of morphological types. Based on our morphological classification, we have found $\sim49\%$ ellipticals (E+dEs), $\sim35\%$ lenticulars and $\sim16\%$ (Spiral + Irr + pec). This is in accordance with the morphological fractions found in high density environments like rich clusters of galaxies. A detailed morphology, their structural parameters are presented along with the catalogue.\\

\noindent {\bf 2.} The velocity dispersion of the cluster is found to be  $\sigma_{cl} \sim 1004$~kms$^{-1}$. This value is in agreement with several other studies in literature.

\noindent {\bf 3.} We present structural parameters for 179 galaxies in the sample using 2D multi-component decomposition in SDSS g, r, i bands. E+dE galaxies are well fit by single Sersic model while S0s and spirals are fitted with double Sersic and exponential depending on whether they host a bar and a bulge. $\sim 75\%$ of the S0s host bulges with Sersic indices $n_{bulge}>2$ - suggesting classical bulge but the remaining might be hosting pseudobulge.\\

\noindent {\bf 4.} Dwarf ellipticals are about 2 mag fainter than their brighter counterpart in our sample and follow a different Sersic index and central velocity dispersion distribution compared to the brighter Es. The median Sersic index for Es and dEs are 4.1 and 2.5 respectively; while the median $\sigma_{vel}$ for the Es and dEs are 164.2 and 61.7 kms$^{-1}$ respectively.\\

\noindent {\bf 5.} Most of the galaxies in the cluster follow a red-sequence in the CMD. Our galaxy sample in the cluster lacks bluer and perhaps, younger population; the CMD is nearly unimodal suggesting strong effects of the cluster environment impacting the galaxy evolution. The cluster ellipticals (E+dE) and lenticulars follow a definite color-magnitude relation with a slope of $-0.057$ over a range of $\sim6$ mag between the fainter and the brighter ones.\\

\noindent {\bf 6.} We have studied KR for our cluster ETGs using 5 different sample construction.
The KR for ETGs based on visual morphology +  central velocity dispersion show the least scatter. The mean slope and zero-point of the KR based on all the five samples are $3.02\pm 0.1$ and $18.65 \pm 0.03$ in the r-band. The KR slope for the ellipticals in our sample matches well with the Coma cluster as well as other clusters at higher redshift. We do not find any systematic variation of the KR slopes across the three SDSS passbands except the zero-point. \\

\noindent \textbf{7.} The dEs in the sample1 prefer a sightly higher KR slope $3.44\pm0.21$ with a zero-point of $20.0\pm 0.05$, about 1.5 mag fainter than Es in r-band. The dEs in Sample2 and Sample3 are also outside $1\sigma$ scatter. In the complimentary Sample5, most dEs and a few S0s are outside $1\sigma$ scatter and they form a parallel KR. \\

\noindent \textbf{8.} We have established the FPR for the cluster normal ellipticals in SDSS g,r and i-bands, with $a = 1.38 \pm 0.003$, $b = 0.33 \pm 0.06$ and $c = -9.18 \pm 0.03$ in r-band. Some of the dEs and S0s with fainter mean surface brightness and low central velocity dispersion from the complimentary sample5 tend to deviate from the linear FPR.


\section*{Acknowledgements}
We thank the anonymous referee for constructive comments and suggestions which have improved the quality of our paper. We are thankful to ISRO RESPOND PROGRAM for supporting us under grant number ISRO/RES/2/417/17-18 to carry out our research work properly. SAK and NI are also thankful to IUCAA Pune and Kashmir University for providing the necessary facilities during the course of this work. The research of IP is supported by the INSPIRE Faculty grant (DST/INSPIRE/04/2016/000404) awarded by the Department of Science and Technology, Government of India.

\section{DATA AVAILABILITY}
We have generated a catalog of 179 galaxies in A426 cluster based on GALFIT analysis. This catalog gives the structural parameters of galaxies in SDSS g, r and i-bands. The catalog will be made available online along with the publication.

\bibliographystyle{mnras}
\bibliography{ref} 




\appendix

\section{SQL QUERY}
\label{sec:query}
\noindent SELECT \\
p.objID, p.ra, p.dec, s.z, s.zErr, s.velDisp, s.velDispErr...\\
FROM photoObj p\\
JOIN dbo.fGetNearbyObjEq(49.946667,41.513056,107) n ON n.objID = p.objID\\
Join specObj s oN s.bestObjID = p.objID\\
JOIN plateX x ON x.plateID = s.plateID\\
where s.class = 'galaxy'\\
and s.zWarning = 0\\

This is an SQl query we used  for searching galaxies in the SDSS database. The \textit{ fGetNearbyObjEq} takes three inputs: right ascension, declination and a radius given in arcminutes. It returns all objects in a circular area centered at the specified right ascension, declination and of radius given above. Through \textit{Join ON} syntax, we join different tables containing photometric and spectroscopic data to obtain the various quantities of interest. The statement \textit{s.class = 'galaxy' }tells the query to return only the galaxies. \textit{s.zWarning = 0} statement ensures that only the galaxies for which redshift is well-defined is returned. Some of the important parameters obtained include ra, dec, z, velocity dispersion, Various types of magnitudes--Petrosian, model magnitudes, effective radii, FWHMs, errors on the measured quantities etc.     

\section{Catalogue description}
We provide a catalog of A426 spectroscopic galaxies in our sample in g,r and i-bands of SDSS. The catalog consists total of 9 tables with three tables corresponding to each g, r and i-bands. In each band, first table consists of structural parameters of  single Sersic fitting, second table consists of structural parameters of (Sersic plus exponential ) fitting and third table consists of structural parameters corresponding to (Double-Sersic plus exponential) fitting.   Table~\ref{tab:s1}, Table~\ref{tab:s2} and Table~\ref{tab:s3} show the column names, descriptions and first 10, 10 and 9 elements of tables in the r-band. We give  below the description of columns in  these tables. The same  description of columns  remain true in g and i bands.Please note that in addition to these 9 tables,there is one more supplementary table
 containing a list of 4 unfitted galaxies in our sample.\\

Description of Table~\ref{tab:s1} \\
-----------------------\\
1\hspace{0.5cm}	$SN$\hspace{1.01cm}	Serial number of a galaxy in our sample.\\
2\hspace{0.5cm}	$RA$\hspace{1cm}	Right ascension in decimal degrees (J2000).\\
3\hspace{0.5cm}	$Dec$	\hspace{0.9cm}Declination in decimal degrees (J2000).\\
4\hspace{0.5cm}	$z$\hspace{1.3cm}	Spectroscopic redshift of a galaxy.\\
5\hspace{0.5cm}	$n$\hspace{1.3cm}	Sersic index of a galaxy.\\
6\hspace{0.5cm}	$m$\hspace{1.2cm}	Apparent r-band magnitude of a galaxy.\\
7\hspace{0.5cm}	$R_e$\hspace{1.1cm}	Effective radius of galaxy in kpc.\\
8\hspace{0.5cm}	$b/a$\hspace{0.9cm}	Apparent axis-ratio of the galaxy.\\
9\hspace{0.55cm}$PA$\hspace{1.1cm}Position angle of galaxy in degrees.\\
10\hspace{0.5cm}$\chi^2$\hspace{1.0cm} Chi-square of fitting.\\
11\hspace{0.4cm}	$T$\hspace{1.2cm}	Morphological type of the galaxy.\\

Description of  Table~\ref{tab:s2} \\

-----------------------\\
1\hspace{0.5cm}	$SN$\hspace{1cm}	Serial number of a galaxy in our sample.\\
2\hspace{0.5cm}	$RA$\hspace{1cm}	Right ascension in decimal degrees (J2000).\\
3\hspace{0.55cm}$Dec$	\hspace{.93cm}Declination in decimal degrees (J2000).\\
4\hspace{0.5cm}	$z$\hspace{1.3cm}	Spectroscopic redshift of a galaxy.\\
5\hspace{0.5cm}	$n_{bulge}$\hspace{.55cm}	Sersic index of  bulge of a galaxy.\\
6\hspace{0.55cm}$m_{bulge}$ \hspace{.45cm} Bulge magnitude of a galaxy in r-band.\\
7\hspace{0.45cm}	${Re}_{bulge}$\hspace{0.4cm}	Effective radius of galaxy bulge  in kpc.\\
8\hspace{0.45cm}	${(b/a)}_{bulge}$\hspace{.07cm}	Apparent axis- ratio of the galaxy bulge.\\
9\hspace{0.55cm}$PA_{bulge}$\hspace{.40cm}Position angle of bulge in degrees.\\
10\hspace{0.43cm}$m_{disk}$\hspace{0.7cm} Disk magnitude of a galaxy in r-band.\\
11\hspace{0.41cm}$R_{s}$\hspace{1.15cm}Scale length of disk in kpc.\\
12\hspace{0.38cm}${(b/a)}_{disk}$\hspace{.17cm} Disk axis-ratio of a galaxy.\\
13\hspace{0.45cm}$PA_{disk}$\hspace{.40cm}Position angle of disk in degrees.\\
14\hspace{0.45cm}$m_{tot}$\hspace{0.75cm} Total magnitude of a galaxy in r-band.\\
15\hspace{0.5cm}$\chi^2$\hspace{1.1cm}Chi-square of fitting.\\
16\hspace{0.45cm}$T$\hspace{1.2cm}	Morphological type of the galaxy.\\

Description of  Table~\ref{tab:s3} \\
-----------------------\\
1\hspace{0.5cm}	$SN$\hspace{1cm}	Serial number of a galaxy in our sample.\\
2\hspace{0.5cm}	$RA$\hspace{1cm}	Right ascension in decimal degrees (J2000).\\
3\hspace{0.5cm} $Dec$	\hspace{0.9cm}Declination in decimal degrees (J2000).\\
4\hspace{0.5cm}	$z$\hspace{1.3cm}	Spectroscopic redshift of a galaxy.\\
5\hspace{0.5cm}	$n_{bulge}$\hspace{.55cm}	Sersic index of  bulge of a galaxy.\\
6\hspace{0.55cm}$m_{bulge}$ \hspace{.47cm} Bulge magnitude of a galaxy in r-band. \\
7\hspace{0.5cm}	${Re}_{bulge}$\hspace{.38cm}	Effective radius of galaxy bulge in kpc.\\
8\hspace{0.5cm}	${(b/a)}_{bulge}$\hspace{.08cm}	Apparent axis-ratio of the galaxy bulge.\\
9\hspace{0.58cm}$PA_{bulge}$\hspace{.40cm}Position angle of bulge in degrees.\\
10\hspace{0.5cm}$n_{bar}$\hspace{.8cm}Sersic index of a bar.\\
11\hspace{0.4cm}$m_{bar}$\hspace{0.8cm}Bar magnitude of a galaxy in r-band.\\
12\hspace{0.5cm}$r_{bar}$\hspace{.92cm}Bar effective radius of a galaxy in kpc.\\
13\hspace{0.5cm}${(b/a)}_{bar}$\hspace{.20cm} Axis-ratio of a bar in galaxy.\\
14\hspace{0.55cm}$PA_{bar}$\hspace{.43cm}Position angle of bar in degrees.\\
15\hspace{0.5cm}$m_{disk}$\hspace{0.7cm}Disk magnitude of a galaxy in r-band.\\
16\hspace{0.5cm}$R_{s}$\hspace{1.1cm}Scale length of disk in kpc.\\
17\hspace{0.5cm}${(b/a)}_{disk}$\hspace{.1cm} Disk axis-ratio of a galaxy.\\
18\hspace{0.55cm}$PA_{disk}$\hspace{.40cm}Position angle of disk in degrees.\\
19\hspace{0.5cm}$m_{tot}$\hspace{0.68cm} Total magnitude of a galaxy in r-band.\\
20\hspace{0.5cm}$\chi^2$\hspace{1cm}Chi-square of fitting.\\
21\hspace{0.5cm}$T$\hspace{1.1cm}	Morphological type of the galaxy.\\

\begin{table*}
	\centering
	\begin{minipage}{171mm}
		\caption{Structural parameters of few  galaxies obtained by fitting single Sersic profile to the images of galaxies in r-band . The full table will be available  online.}
		\scriptsize
		\label{tab:s1}
		\begin{tabular}{@{}ccccccccccl@{}}
			\hline \hline
			 $SN$ & $RA$ & $Dec$ & $z$ &  $n$ & $m$ & $R_e$ & $b/a$ & $PA$ & $\chi^2$ & $T$ \\
			 (1) & (2) & (3) & (4) & (5) & (6) & (7) & (8) & (9) & (10) & (11)\\
			\hline

3 & 50.1375 & 41.5741 & 0.0106 & 4.21 & 14.36 & 0.45 & 0.61 & 86.98 & 1.206 & cE(bright) \\ 
8 & 49.5789 & 41.4694 & 0.0118 & 3.44 & 14.4 & 1.21 & 0.48 & -73.99 & 1.109 & dE \\ 
10 & 50.8608 & 42.0641 & 0.0121 & 2.6 & 15.1 & 1.07 & 0.49 & -78.6 & 1.033 & dE/E5 \\ 
12 & 50.1758 & 41.404 & 0.0125 & 5.97 & 13.98 & 2.88 & 0.58 & -11.25 & 1.819 & E4 \\ 
14 & 51.1519 & 40.6908 & 0.0126 & 5.33 & 12.32 & 4.91 & 0.81 & -40.13 & 0.818 & gE \\ 
16 & 50.3803 & 41.5475 & 0.0129 & 3.48 & 15.95 & 0.52 & 0.66 & 23.54 & 1.107 & dE \\ 
18 & 49.9326 & 41.457 & 0.013 & 5.08 & 14.58 & 1.16 & 0.93 & -25.98 & 1.178 & cE \\ 
19 & 50.1047 & 42.5471 & 0.0133 & 4 & 12.57 & 5.52 & 0.76 & 85.51 & 1.657 & E2 \\ 
21 & 50.2558 & 41.4345 & 0.0134 & 5.89 & 15.6 & 0.54 & 0.64 & 21.55 & 1.11 & cE \\ 
22 & 48.9153 & 41.1469 & 0.0135 & 1.79 & 15.36 & 2.02 & 0.74 & -14.55 & 1.018 & dE \\ 

				\hline
		\end{tabular}
		
	\end{minipage}
	\normalsize
\end{table*}

\begin{figure}
\includegraphics[width=0.45\textwidth]{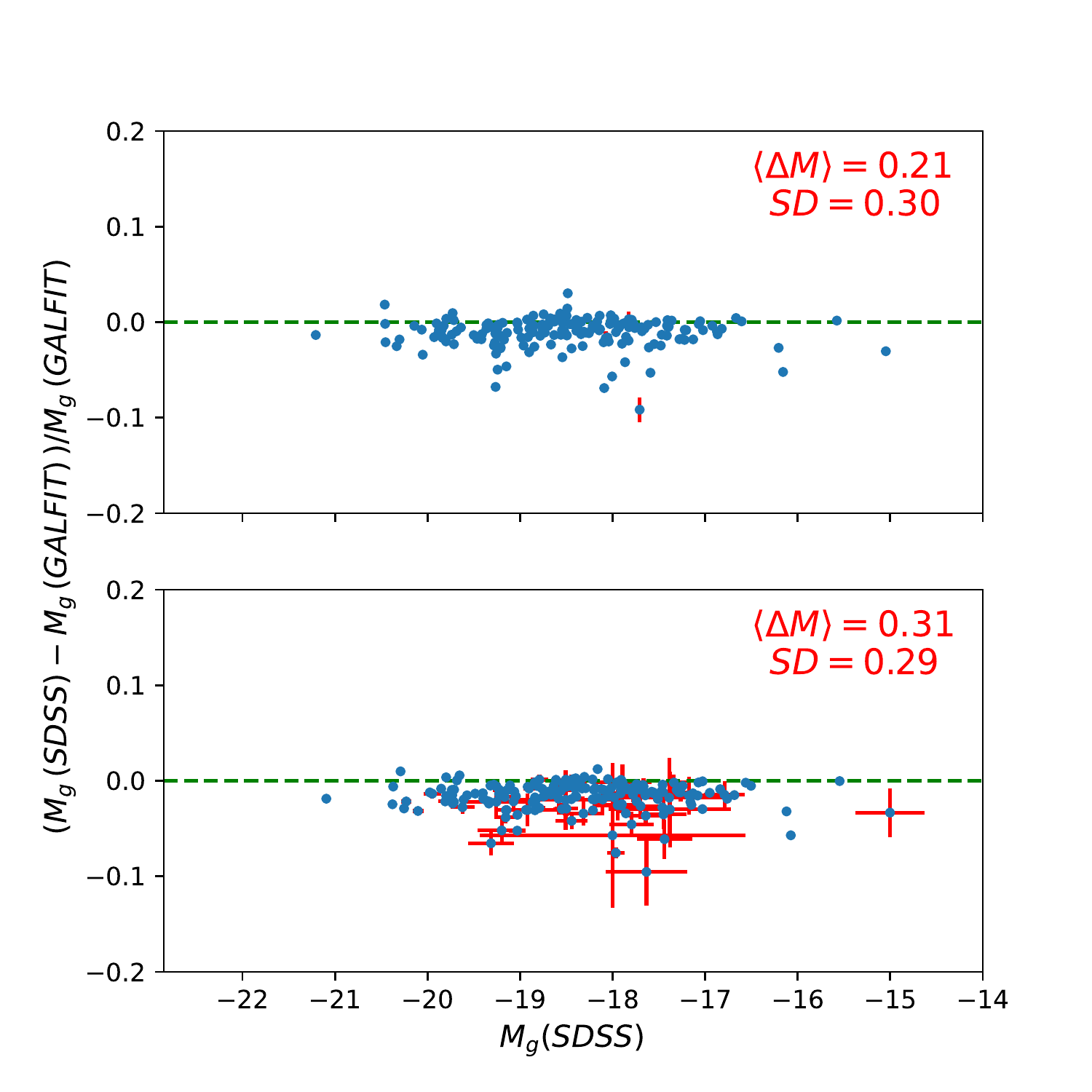}
\includegraphics[width=0.45\textwidth]{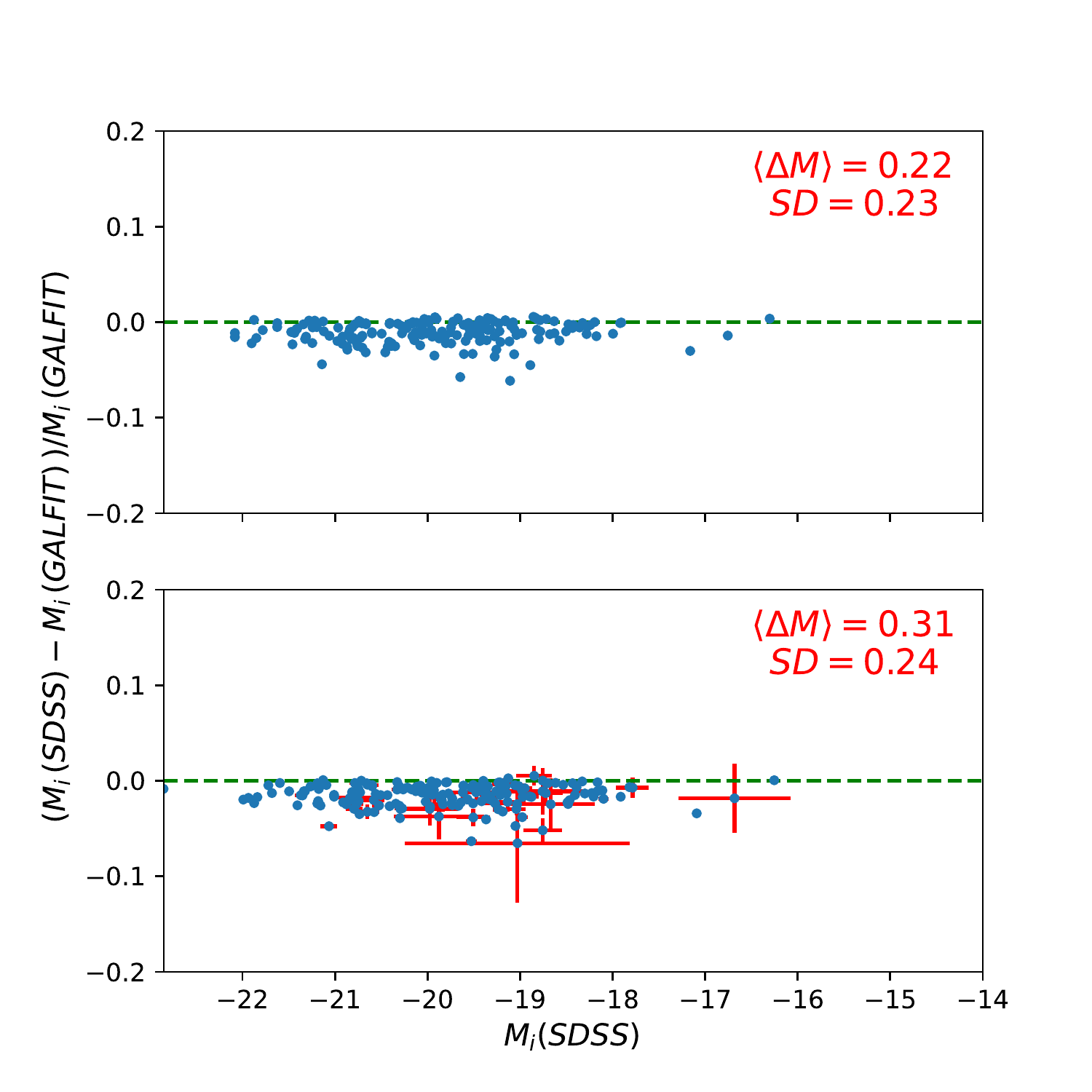}
\caption{Comparison of absolute magnitudes in g-band (upper panels) and i band (bottom panels).}
\label{fig:gidis}
\end{figure}

\begin{table*}
	\centering
	\begin{minipage}{171mm}
		\caption{Structural parameters of few   galaxies obtained by fitting  Sersic plus Exponential   profile to the images of galaxies in  r-band. The full table will be available  online.}
		\scriptsize
		\label{tab:s2}
		\begin{tabular}{@{}cccccccccccccccl@{}}
			\hline \hline
			$SN$ & $RA$ &
			$Dec$ & $z$  & $n_{bulge}$ & $m_{bulge}$ & $Re_{bulge}$ & $(b/a)_{bulge}$ & $PA_{bulge}$& $m_{disk}$ & $R_s$ & $(b/a)_{disk}$ & $PA_{disk}$ & $m_{tot}$ & $\chi^2$ & $T$ \\
			(1) & (2) & (3) & (4) & (5) & (6) & (7) & (8) & (9) & (10) & (11) & (12) & (13) & (14) &(15) & (16)  \\
			\hline

13 & 49.46 & 41.9676 & 0.0126 & 1.28 & 16.22 & 0.55 & 0.57 & 79.25 & 13.74 & 2.3 & 0.17 & -34.73 & 13.63 & 1.338 & Sa \\ 
30 & 49.5073 & 41.5927 & 0.014 & 3.58 & 15.3 & 0.79 & 0.79 & -0.78 & 15.41 & 1.87 & 0.32 & 4.02 & 14.6 & 1.378 & S0 \\ 
31 & 50.0216 & 40.9062 & 0.014 & 4.78 & 14.3 & 1.46 & 0.52 & 48.55 & 15.16 & 1.69 & 0.19 & 53.72 & 13.89 & 1.147 & S0 \\ 
49 & 48.5351 & 41.2921 & 0.0152 & 2.81 & 14.27 & 0.62 & 0.92 & -44.53 & 13.59 & 4.49 & 0.47 & 78.61 & 13.12 & 1.416 & S0lens \\ 
59 & 50.2517 & 41.5625 & 0.0157 & 2.11 & 14.6 & 0.36 & 0.75 & -56.22 & 14.54 & 1.47 & 0.35 & -48.78 & 13.82 & 1.269 & S0 \\ 
64 & 49.43 & 42.3851 & 0.0161 & 3.49 & 14.53 & 1.81 & 0.58 & 63.89 & 13.93 & 3.05 & 0.33 & 43.3 & 13.44 & 2.633 & S0+ \\ 
66 & 50.2869 & 41.253 & 0.0162 & 0.81 & 17.21 & 0.24 & 0.63 & -18.6 & 15.41 & 1.22 & 0.45 & -22.07 & 15.22 & 1.098 & S0a \\ 
72 & 49.191 & 40.33 & 0.0164 & 0.36 & 18.07 & 0.19 & 0.87 & -20.37 & 14.43 & 1.67 & 0.31 & -37.04 & 14.39 & 1.622 & Sc \\ 
78 & 49.3067 & 41.4355 & 0.0166 & 1.96 & 15.78 & 0.32 & 0.74 & 88.97 & 15 & 1.3 & 0.84 & 84.21 & 14.57 & 1.064 & SB0 \\ 
90 & 50.8952 & 40.953 & 0.0172 & 2.88 & 16.46 & 0.78 & 0.61 & -4.96 & 15.49 & 1.84 & 0.23 & -16.39 & 15.12 & 1.047 & Sc \\

         \hline
		\end{tabular}
		
	\end{minipage}
	\normalsize
\end{table*}

\begin{table*}
	\centering
	\begin{minipage}{175mm}
		\caption{Structural parameters of few   galaxies obtained by fitting double- Sersic plus Exponetial profile to the images of galaxies in r-band. The full table will be available  online.}
		\scriptsize
		\label{tab:s3}
		\begin{tabular}{@{}cccccccccccccccl@{}}
			\hline \hline
			$SN$ &$RA$ & $Dec$ & $z$  & $n_{bulge}$ & $m_{bulge}$ & $Re_{bulge}$ & $(b/a)_{bulge}$ & $PA_{bulge}$ & $n_{bar}$ & $m_{bar}$ & $r_{bar}$ & $(b/a)_{bar}$ & $PA_{bar}$& $m_{disk}$ & $R_s$  \\
			
			(1) & (2) & (3) & (4) & (5) & (6) & (7) & (8) & (9) & (10) & (11) & (12) & (13) & (14) & (15) & (16) \\
			\hline

61 & 50.5484 & 40.7268 & 0.0159 & 4.42 & 14.79 & 0.82 & 0.7 & 22.9 & 0.92 & 14.46 & 4.61 & 0.63 & 17.83 & 14.46 & 2.87 \\ 
156 & 50.5378 & 41.4508 & 0.0209 & 3.3 & 16.15 & 0.93 & 0.94 & -81.79 & 0.59 & 16.68 & 2.33 & 0.35 & -55.24 & 15.37 & 3.44 \\ 
6 & 49.4982 & 41.5204 & 0.0111 & 2.92 & 14.21 & 0.76 & 0.82 & -87.37 & 0.43 & 14.82 & 2.21 & 0.25 & -81.86 & 12.86 & 3.86 \\ 
81 & 50.2593 & 41.4095 & 0.0167 & 1.92 & 15.12 & 0.58 & 0.91 & 40.11 & 0.58 & 15.99 & 2.25 & 0.33 & 47.49 & 14.48 & 3.26 \\ 
93 & 49.697 & 41.9806 & 0.0175 & 2.61 & 16.41 & 1.08 & 0.59 & 42.04 & 0.08 & 18.25 & 1.77 & 0.52 & 5.21 & 15.35 & 3.34 \\ 
98 & 50.9535 & 40.5578 & 0.0178 & 2.72 & 14.21 & 0.9 & 0.92 & -50.84 & 0.18 & 15.52 & 4.93 & 0.3 & -13.03 & 12.89 & 5.06 \\ 
170 & 50.1725 & 42.8041 & 0.0227 & 4 & 17.19 & 0.18 & 0.71 & -85.14 & 0.69 & 15.97 & 4.76 & 0.21 & -68.95 & 14.2 & 4.54 \\ 
58 & 49.0032 & 40.8857 & 0.0157 & 0.9 & 17.64 & 0.19 & 0.82 & 70.51 & 0.34 & 16.79 & 1.64 & 0.31 & 43.35 & 13.03 & 6.28 \\ 
41 & 50.4324 & 42.351 & 0.0144 & 3.12 & 16.68 & 0.17 & 0.73 & 50.52 & 0.55 & 16.61 & 1.46 & 0.28 & 78.48 & 14.49 & 1.83 \\ 
\hline
\end{tabular}
\end{minipage}
\normalsize
\begin{minipage}{175mm}
\scriptsize
\begin{tabular}{@{}ccccl@{}}
\hline \hline
$(b/a)_{disk}$ & $PA_{disk}$& $m_{tot}$& $\chi^2$ & $T$ \\
(17) & (18) & (19) & (20) & (21)\\
\hline
0.57 & 23.06 & 13.36 & 1.343 & SBa(r) \\ 
0.87 & 4.01 & 14.74 & 1.131 & SB0 \\ 
0.93 & -25.42 & 12.45 & 1.09 & SB0 \\ 
0.94 & 16.07 & 13.84 & 1.143 & SB0 \\ 
0.51 & -12.87 & 14.95 & 0.598 & SB0 \\ 
0.94 & 24.64 & 12.53 & 2.012 & SB0lens \\ 
0.82 & -62.15 & 13.95 & 1.063 & SB0 \\ 
0.39 & 59.25 & 12.98 & 1.337 & SABd \\ 
0.93 & 55.67 & 14.23 & 1.185 & SBb \\ 
\hline
\end{tabular}
\end{minipage}
\normalsize
\end{table*}
	
\section{Absolute magnitude plots}
\label{sec:Abs_gi}
A comparison of absolute magnitudes obtained through GALFIT and SDSS in g and i bands are shown in Fig.~\ref{fig:gidis}. 

SDSS provides several "Model" magnitudes for each object, e.g., de Vaucouleurs magnitude(devMag) and exponential magnitude(expMag). These are obtained by independently fitting a pure de Vaucouleurs profile and a pure exponential profile respectively to the two-dimensional image of each object in each band. Note that de Vaucouleurs profile is truncated beyond $7r_e$ to smoothly go to zero at $8r_e$, and with some softening within $r=r_e/50$ and exponential profile is  truncated beyond $3r_e$ to smoothly go to zero at $4r_e$. Between these two models, the model of higher likelihood is chosen in r-band and applied in the other bands after convolving with appropriate PSF in each band. The resulting magnitudes are termed as modelMag.
For the calculation of Composite Model Magnitudes( cModelMag), the best fit exponential and de Vaucouleurs fits in each band is taken and the linear combination of the two that best fits the image is done. From the coefficient(fracDeV) of the de Vaucouleurs term, composite flux is calculated using

\begin{equation}
    F_{composite} = fracDeV \times F_{deV} + (1 - fracDeV)\times F_{exp},
\end{equation}

where $F_{deV}$ and $F_{exp}$ are the de Vaucouleurs and exponential fluxes of the object. The magnitude obtained from $F_{composite}$ is termed as the cmodel magnitude. SDSS uses a modified form of \cite{Petrosian_1976} system to measure galaxy fluxes within a circular aperture whose radius is defined by the shape of the azimuthally averaged light profile \citep{Blanton_2001,Yasuda_2001} and the magnitude with the petrosian radius is defined as the Petrosian magnitude.

\bsp	
\label{lastpage}
\end{document}